\documentclass{article}

\usepackage{amsmath,amsfonts,bm}

\def\eqref#1{equation~\ref{#1}}

\def\1{\bm{1}}

\def\rmI{{\mathbf{I}}}

\def\rmM{{\mathbf{M}}}

\def\rmW{{\mathbf{W}}}
\def\rmX{{\mathbf{X}}}

\def\ermM{{\textnormal{M}}}

\def\ermX{{\textnormal{X}}}

\def\estpx{{\widehat{p}_\rmX}}

\def\mMc{{\mM^{\text{cond}}}}

\def\rmMc{{\rmM^{\text{cond}}}}
\def\idK{{\mathcal{K}}}
\def\emMc{{\emM^{\text{cond}}}}

\def\idC{{\mathcal{C}}}

\def\idR{{\mathcal{R}}}

\def\vep{{\bm{\varepsilon}}}

\def\vone{{\bm{1}}}

\def\vh{{\bm{h}}}

\def\vl{{\bm{l}}}

\def\vs{{\bm{s}}}

\def\mI{{\bm{I}}}

\def\mM{{\bm{M}}}

\def\mX{{\bm{X}}}
\def\mY{{\bm{Y}}}

\DeclareMathAlphabet{\mathsfit}{\encodingdefault}{\sfdefault}{m}{sl}
\SetMathAlphabet{\mathsfit}{bold}{\encodingdefault}{\sfdefault}{bx}{n}

\def\gB{{\mathcal{B}}}
\def\gC{{\mathcal{C}}}

\def\gR{{\mathcal{R}}}

\def\sC{{\mathbb{C}}}

\def\sM{{\mathbb{M}}}

\def\sR{{\mathbb{R}}}

\def\sW{{\mathbb{W}}}

\def\emM{{M}}

\def\emX{{X}}

\newcommand{\E}{\mathbb{E}}

\newcommand{\R}{\mathbb{R}}

\DeclareMathOperator*{\argmin}{arg\,min}

\usepackage{adjustbox}
\usepackage{musicography}
\usepackage{multirow}
\usepackage[linesnumbered,ruled,vlined]{algorithm2e}
\usepackage{url}
\usepackage{stmaryrd}
\usepackage{subcaption}
\usepackage{mathrsfs}
\usepackage{graphicx}
\usepackage{amsthm}
\usepackage{amssymb, amsmath}

\usepackage{colortbl}  
\usepackage[table,xcdraw]{xcolor}    
\usepackage{booktabs}  
\usepackage{graphicx}  

\usepackage{hyperref}

 \usepackage[accepted]{icml2025}

\usepackage{mathtools}

\usepackage[capitalize,noabbrev]{cleveref}

\newtheorem{proposition}{Proposition}
\newtheorem{remark}{Remark}
\newtheorem{lemma}{Lemma}

\newtheorem{definition}{Definition}

\usepackage[textsize=tiny]{todonotes}

\icmltitlerunning{Efficient Fine-Grained Guidance for Diffusion Model Based Symbolic Music Generation}

\begin{document}

\twocolumn[
\icmltitle{Efficient Fine-Grained Guidance for Diffusion Model Based Symbolic Music Generation}



\icmlsetsymbol{equal}{*}

\begin{icmlauthorlist}
\icmlauthor{Tingyu Zhu}{equal,yyy}
\icmlauthor{Haoyu Liu}{equal,yyy}
\icmlauthor{Ziyu Wang}{comp}
\icmlauthor{Zhimin Jiang}{xxx}
\icmlauthor{Zeyu Zheng}{yyy}
\end{icmlauthorlist}

\icmlaffiliation{yyy}{University of California, Berkeley, USA}
\icmlaffiliation{comp}{New York University, New York, USA}
\icmlaffiliation{xxx}{Touka Technologies}

\icmlcorrespondingauthor{Haoyu Liu}{haoyuliu@berkeley.edu}
\icmlcorrespondingauthor{Zeyu Zheng}{zyzheng@berkeley.edu}

\icmlkeywords{Machine Learning, ICML}

\vskip 0.3in
]



\printAffiliationsAndNotice{\icmlEqualContribution} 

\begin{abstract}
Developing generative models to create or conditionally create symbolic music presents unique challenges due to the combination of limited data availability and the need for high precision in note pitch. To address these challenges, we introduce an efficient Fine-Grained Guidance (FGG) approach within diffusion models. FGG guides the diffusion models to generate music that aligns more closely with the control and intent of expert composers, which is critical to improve the accuracy, listenability, and quality of generated music. This approach empowers diffusion models to excel in advanced applications such as improvisation, and interactive music creation. We derive theoretical characterizations for both the challenges in symbolic music generation and the effects of the FGG approach. We provide numerical experiments and subjective evaluation to demonstrate the effectiveness of our approach. We have published a demo page \footnote{The demo page is available at \href{https://huajianduzhuo-code.github.io/FGG-diffusion-music/}{https://huajianduzhuo-code.github.io/FGG-diffusion-music/}, we also release the complete source code at \href{https://github.com/huajianduzhuo-code/FGG-music-code}{https://github.com/huajianduzhuo-code/FGG-music-code}} to showcase performances, which enables real-time interactive generation. 
\end{abstract}

\section{Introduction}
\label{sec:intro}


Diffusion models \cite{ho2020denoising} have consistently demonstrated effectiveness across a wide range of generative tasks, particularly in image and video generation \cite{rombach2022high}. Despite success, diffusion models face some limitations. (1) Imprecise detail generation: Diffusion models often struggle with accurately producing details, leading to artifacts or distortions in the generated content, such as noticeable inconsistencies or distortions in videos.
(2) Limited controllability: Obtaining precise control over the generated content to align it with the intent of the user remains a significant challenge. For instance, correcting specific distortions in a generated video while keeping the rest of the scene unchanged is difficult with current diffusion model frameworks. 

These limitations are exacerbated in situations where data is scarce, which is often the case in domains like symbolic music generation, where symbolic music data is
limited due to copyright constraints and the effort needed to create data. Additionally, unlike image generation, where the inaccuracy of a single pixel may not significantly affect overall quality, symbolic music generation
demands high precision, especially in terms of pitch. In many musical and tonal contexts, even a single incorrect or inconsistent note can be glaringly obvious and disturbing.


To provide more contexts, symbolic music generation is a subfield of music generation that focuses on creating music in symbolic form, typically represented as sequences of discrete events such as notes, pitches, rhythms, and durations. These representations are analogous to traditional sheet music or MIDI files, where the structure of the music is defined by explicit musical symbols rather than audio waveforms.  Many recent works in symbolic music generation are based on diffusion models; see \citet{min2023polyffusion}, \citet{wang2024whole} and  \citet{huang2024symbolic} for example. 

Following this branch of work, we address the precision and controllability challenges in diffusion-based symbolic music generation by incorporating fine-grained guidance into the training and sampling processes. While soft control schemes such as providing chord conditions may fail to ensure detailed pitch correctness, we propose to enhance chord conditioning with a hard control method integrated into the sampling process, which guarantees the desired tonal correctness in every generated sample.

Our results in this work are summarized as follows:
\begin{itemize}
    \item \textbf{Motivation}: We theoretically and empirically characterize the challenge of precision in symbolic music generation

    \item \textbf{Methodology}: We incorporate fine-grained harmonic and rhythmic guidance to symbolic music generation with diffusion models.

    \item \textbf{Functionality}: The developed model is capable of generating music with high accuracy in pitch and consistent rhythmic patterns that align closely with the user’s intent.

    \item \textbf{Effectiveness}: We provide both theoretical and empirical evidence supporting the effectiveness of our approach.
\end{itemize}

\subsection{Related Work}
\label{sec:related_work}

\paragraph{Symbolic Music Generation.}
Symbolic music generation literature can be classified based on the choice of data representation, among which the
MIDI token-based representation adopts a sequential discrete data structure, and is often combined with sequential generative models such as Transformers and LSTMs. 

To leverage well-developed generative models for symbolic music, \citet{huang2018music} introduced a Transformer-based model with a novel relative attention mechanism designed for symbolic music generation. Subsequent works have enhanced the controllability of symbolic music generation by incorporating input conditions. For instance, \citet{huang2020pop} integrated metrical structures to enhance rhythmic coherence, \citet{ren2020popmag} conditioned on melody and chord progressions for harmonically guided compositions, and \citet{choi2020encoding} encoded musical style to achieve nuanced harmonic control. These advancements have contributed to more interpretable and user-directed music generation control.

To better capture spatio-temporal harmonic structures in music, researchers have adopted diffusion models with various control mechanisms. \citet{min2023polyffusion} incorporated control signals tailored to diffusion inputs, enabling control over melody, chords, and texture. \citet{wang2024whole} extended this by integrating hierarchical control for full-song generation. To further enhance control, \citet{zhang2023sdmuse} and \citet{huang2024symbolic} leveraged the gradual denoising process to refine sampling. Building on these approaches, our work addresses the remaining challenge of precise control in real-time generation.

In parallel to diffusion-based approaches, a body of work on general symbolic music generation—using models such as RNNs, GANs, and VAEs—has also explored mechanisms for achieving precise user-controllable generation. Early work on symbolic generation already explored user-steerable conditioning. \citet{meade2019exploring} retrofitted an RNN method with human-interpretable controls such as note density and pitch range limits. \citet{Dong2017MuseGANMS} proposed a method that conditions a GAN model on one track given by human to generate the remaining tracks based on temporal structure of that track. \citet{Wu2021MuseMorphoseFA} utilized a Transformer based VAE model to realize fine-grained style transfer over full songs.

\paragraph{Image Inpainting.}
Image inpainting with diffusion models has advanced rapidly, offering valuable insights for our task. In our setting, harmonic conditions define constrained or masked regions, and the model must complete the rest—analogous to inpainting. Recent diffusion-based methods have enabled fine-grained control during both training and sampling. For instance, \citet{Lugmayr2022RePaintIU} introduced a post-conditioning strategy that adapts the reverse diffusion process to reconcile known and missing regions without retraining, albeit with increased inference time. \citet{Xie2022SmartBrushTA} combined shape and text prompts to enable precise, user-guided inpainting via joint training and sampling design. \citet{Corneanu2024LatentPaintII} improved sampling efficiency by conditioning directly in the latent space, supporting faster and semantically coherent completions. Inspired by these works, we adapt the idea of context-aware, guided completion to symbolic music, enabling controllable generation over structured time-pitch domains.


\paragraph{Controlled Diffusion Models.}

Multiple works in controlled diffusion models are related to our work in terms of methodology.
Specifically, we adopt the idea of  classifier-free guidance in training and generation, see \citet{ho2021classifier}. To control the sampling process, \citet{chung2022diffusion}, \citet{song2023loss} and \citet{novack2024ditto} guide the intermediate sampling steps using the gradients of a loss function. In contrast, \citet{dhariwal2021diffusion},  \citet{saharia2022photorealistic}, \citet{lou2023reflected} and \citet{fishman2023diffusion} apply projection and reflection during the sampling process to straightforwardly incorporate data constraints. 
Different from these works, we design guidance for intermediate steps tailored to the unique characteristics of symbolic music data and generation. While the meaning of a specific pixel in an image is undefined until the entire image is generated, each position on a piano roll corresponds to a fixed time-pitch pair from the outset. This new context enables us to develop novel implementations and theoretical perspectives on the guidance approach.

\section{Background: Diffusion Models for Piano Roll Generation}
\label{sec:background}

In this section, we introduce the data representation of piano roll. We then introduce the formulations of diffusion model, combined with an application on modeling the piano roll data. 

\paragraph{Data Representation of Piano Rolls.} Let $\rmM\in\{0,1\}^{L\times H}$ be a piano roll segment, where $H$ is the pitch range and $L$ is the number of time units in a frame. For example, $H$ can be set as $128$, representing a pitch range of $0-127$, and $L$  as $64$, representing a 4-bar segment with time signature 4/4 (4 beats per bar) and 16th-note resolution. Each element $\ermM_{lh}$ of $\rmM$ $(l\in\llbracket 1, L\rrbracket ,~h\in\llbracket 1, H\rrbracket)$ takes value $0$ or $1$, where $\ermM_{lh}=1/0$ represents the presence/absence of a note at time index $l$ and pitch $h$.\footnote{This is a slightly simplified representation model for the purpose of theoretical analysis, the specified version with implementation details is provided in Section \ref{subsec:experiment} } Since standard diffusion models are based on Gaussian noise, the output of the diffusion model is a continuous random matrix $\rmX\in\R^{L\times H}$, which is then projected to the discrete piano roll $\rmM$ by
$\ermM_{lh}(\rmX) = \vone\{\ermX_{lh}\geq 1/2\}$, where $\vone\{\cdot\}$ stands for the
indicator function.

\paragraph{Formulation of the Diffusion Model.} To model and generate the distribution of $\rmM$, denoted as $P_{\rmM}$, we use the the Denoising Diffusion Probabilistic Modeling (DDPM) formulation \citep{ho2020denoising}. The objective of DDPM training, with specific choices of parameters and reparameterizations, is given as
\begin{equation}
\label{eq:loss function}    \mathbb{E}_{t\sim\mathcal{U}\llbracket 1,T\rrbracket,\rmX_0\sim P_\rmM,{\vep}\sim\mathcal{N}(0,\mathbf{I})}[\lambda(t)\Vert {\vep}-{\vep}_\theta(\rmX_t,t)\Vert^2],
\end{equation}
where ${\vep}_\theta$ is a deep neural network with parameter $\theta$.
Moreover, according to the connection between diffusion models and score matching \citep{song2019generative}, the deep neural network $\vep_\theta$ can be used to derive an estimator of the score function $\vs_t(\mathbf{X}_t)=\nabla_{\rmX_t}\log p_t(\rmX_t)$. Specifically, $\vs_t(\mathbf{X}_t)$ can be approximated by $-\vep_\theta(\rmX_t,t)/\sqrt{1-\bar{\alpha}_t}$.

With the trained noise prediction network $\vep_\theta$, the reverse sampling process can be formulated as \citep{song2020denoising}:
\begin{equation}
\label{eq: DDIM}
\begin{aligned}
    \rmX_{t-1}=\sqrt{\bar{\alpha}_{t-1}}\left(\frac{\rmX_t-\sqrt{1-\bar{\alpha}_t}\vep_\theta(\rmX_t,t)}{\sqrt{\bar{\alpha}_t}}\right) \\
    +\sqrt{1-\bar{\alpha}_{t-1}-\sigma_t^2}\vep_\theta(\rmX_t,t)+\sigma_t\vep_t,
\end{aligned}
\end{equation}
where $\sigma_t$ are hyperparameters chosen corresponding to \eqref{eq:loss function}, and $\vep_t$ is standard Gaussian noise at each step. Going backward in time from $\rmX_T\sim\mathcal{N}(0,\rmI)$, the process yields the final output $\rmX_0$, which can be converted into a piano roll $\rmM(\rmX_0)$.

According to \citet{song2020score}, the DDPM forward and backward processes can be regarded as discretizations of the following SDEs:
\begin{align}
    &d\rmX_t = -\frac{1}{2}\beta(t)\rmX_t dt+\sqrt{\beta(t)} d\rmW_t,\label{eq:forward SDE}\\
    &d\rmX_t = -\left[\frac{1}{2}\beta(t)\rmX_t+\beta(t)\vs_t(\mathbf{X}_t)\right]dt+\sqrt{\beta(t)}d\bar{\rmW}_t,\label{eq: reverse SDE}
    \end{align}

\section{Methodology: Fine-Grained Guidance}
\label{sec:methodology}

While generative models have achieved significant success in text, image, and audio generation, the effective modeling and generation of symbolic music remains a relatively unexplored area. One challenge of symbolic music generation involves the high-precision requirement in harmony. Unlike image generation, where a slightly misplaced pixel
may not significantly affect the overall image quality, an
“inaccurately” generated musical note can drastically disrupt the harmony, affecting the quality of a piece.

In this section, we present a control methodology that can precisely achieve the desired harmony.  Specifically, we design a fine-grained conditioning and sampling control, altogether referred to as \textit{Fine-Grained Guidance} (FGG) that leverage the characteristic of the piano roll data.



\subsection{Fine-Grained Conditioning in Training}
\label{subsec:fine-grained condition}
We first introduce fine-grained conditioning in training, which serves as the foundation of the more important sampling control in the next subsection \ref{subsec:fine-grained sampling}.

We train a conditional diffusion model with fine-grained harmonic ($\gC$, required) and rhythmic ($\gR$, optional) conditions, which are provided to the diffusion models in the form of a piano roll $\mMc$. We provide illustration of $\mMc(\idC,\idR)$ and $\mMc(\idC)$ via examples in Figure~\ref{fig:McondCR} and Figure~\ref{fig:McondC}, respectively. The mathematical descriptions are provided in Appendix \ref{appendix:additional}.
\begin{figure}[ht]
    \centering
    \includegraphics[width=0.7\linewidth]{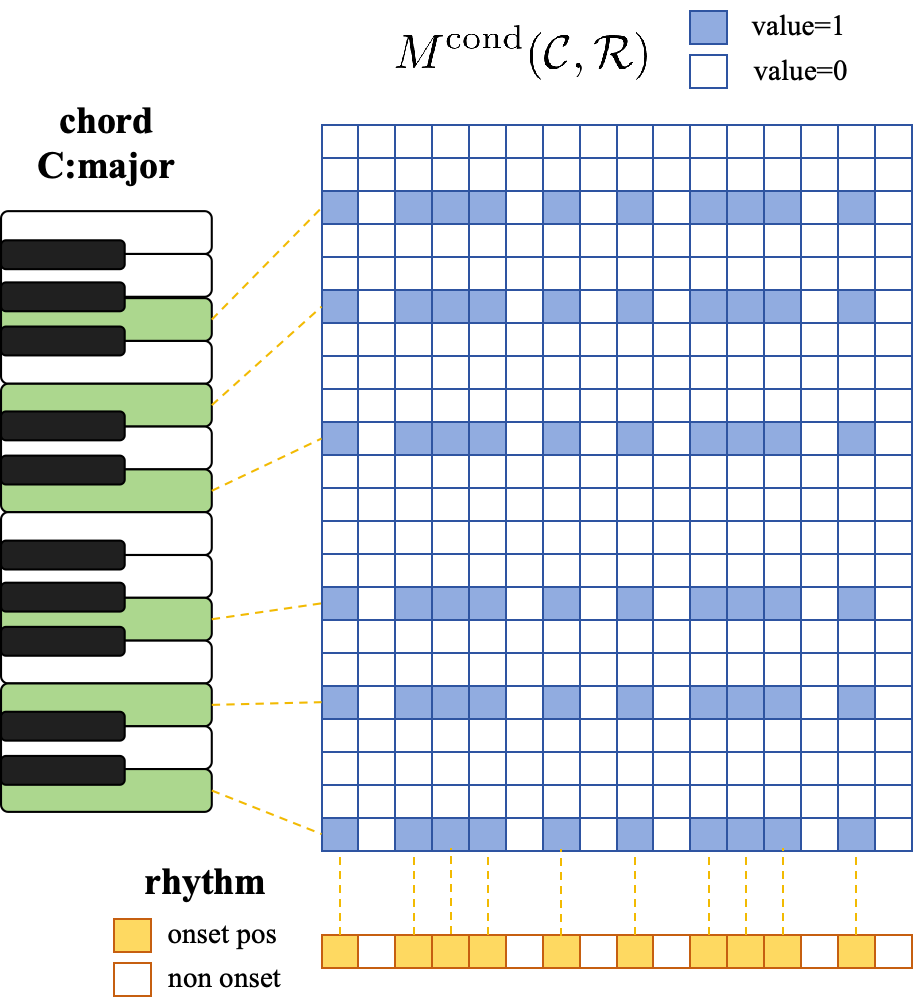}
    \caption{An illustrative example of $\mMc(\idC,\idR)$ with both conditions.}
    \label{fig:McondCR}
\end{figure}

\begin{figure}[ht]
    \centering
    \includegraphics[width=0.7\linewidth]{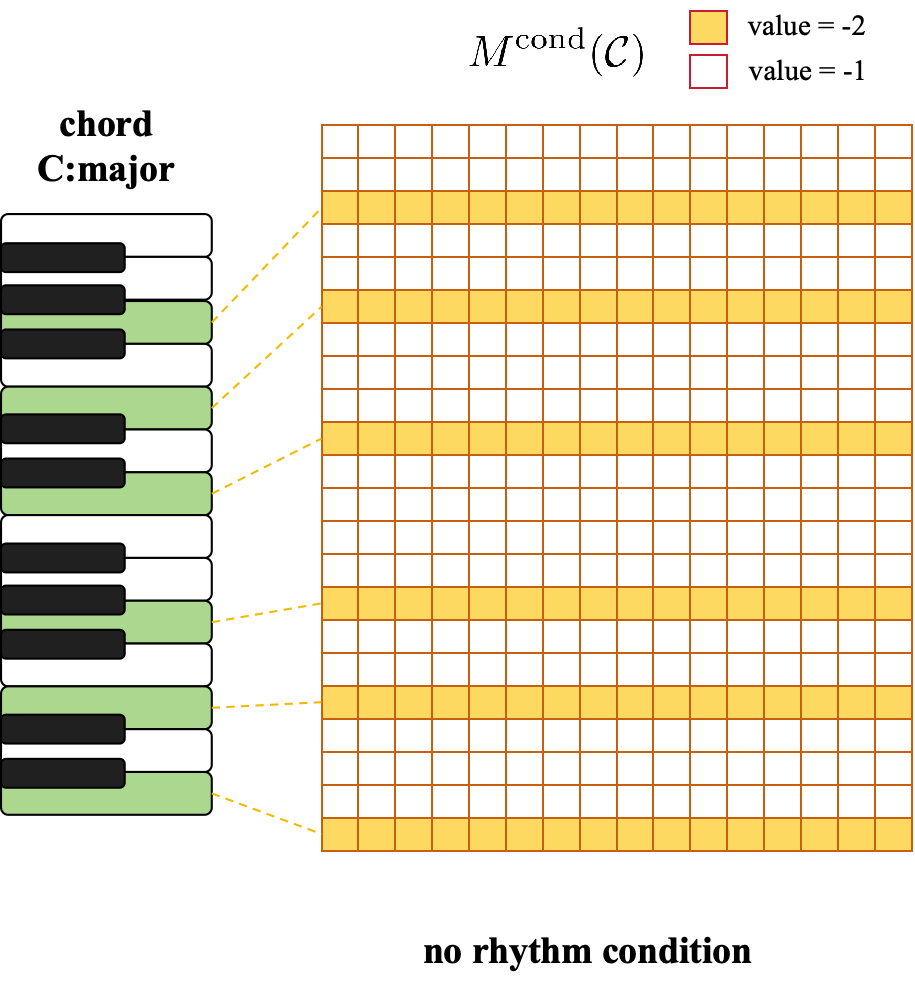}
    \caption{An illustrative example of $\mMc(\idC)$ with harmonic conditions only.\footnotemark}
    \label{fig:McondC}
\end{figure}

\footnotetext{To enable the model to handle both rhythm+chord and chord-only conditions, we use negative values to indicate the absence of rhythmic input when only harmonic conditions are provided, avoiding misinterpretation of $0$s and $1$s as active constraints. Empirically, removing this distinction (i.e., still using $0$ and $1$ when rhythmic condition is not provided) led to a 8–15\% drop in chord accuracy (e.g., direct chord accuracy drops from 0.485 to 0.421, and chord similarity drops from 0.767 to 0.705), highlighting the importance of explicitly encoding missing rhythmic information.}

\subsection{Fine-Grained Control in Sampling Process}
\label{subsec:fine-grained sampling}

We first provide a rough idea of the harmonic sampling control. To integrate harmonic constraints into our model, we employ temporary tonic key\footnote{As a clarification, instead of assigning one single key to a piece or a big section, here we refer to each key associated with the \textit{temporary tonic}.} signatures to establish the tonal center. Our sampling control mechanism guides the gradual denoising process to ensure that the final generated notes remain within a specified set of pitch classes. This control mechanism removes or replaces harmonically conflicting notes, maintaining alignment with the temporary tonic key.


\paragraph{Preliminaries.} Recall that a piano roll segment $\rmM\in\{0,1\}^{L\times H}$, where $l\in\llbracket 1,L\rrbracket$ is the time index, and $h\in\llbracket 1, H\rrbracket$ is the pitch index. For given chord condition sequence $\idC$, let $\idK$ denote the corresponding key sequence. For example, when the C major chord appears as the chord condition at time index $l$, we would expect $\idK(l)$ to contain the pitch classes of the C major scale\footnote{We note that the correspondence between $\mathcal{C}$ and $\mathcal{K}$ is in fact flexible, and can be designed by the user of the model. More discussion is provided in the next section \ref{sec:challenges}}. We note that $\mathcal{C}$ is essentially different from $\mathcal{K}$, where $\mathcal{C}$ describes chord sequences and is provided as condition for generation, and $\mathcal{K}$ is a restriction of ``allowed'' pitch classes for sampling refinement.

Let $w(\vl;\idK):=\{l,w(l;\idK)\}_{l=1}^L$denote the undesired pitch positions on the piano roll $\rmM$. The generated piano roll $\widehat{\rmM}$ is expected to satisfy $\widehat{\ermM}_{lh}=0$, for all $(l,h)\in w(\vl,\idK)$. In other words, for $\widehat{\rmX}_0$ we need 
\begin{equation}
    \forall (l,h)\in w(\vl,\idK), ~P\left(\widehat{\ermX}_{0,lh}> 1/2\right)=0.
    \label{eq:prob0}
\end{equation}
Note that in the backward sampling \eqref{eq: DDIM} that derives $\mX_{t-1}$ from $\mX_t$, we have for the first term \citep{song2020denoising, chung2022diffusion}
\begin{equation}
\begin{aligned}
    &\left(\frac{\mX_t-\sqrt{1-\bar{\alpha}_t}\widehat{\vep}_\theta(\mX_t,t)}{\sqrt{\bar{\alpha}_t}}\right)
    =\text{``predicted $\rmX_0$''}\\&=\widehat{\E}[\rmX_0|\mX_t],\quad t=T,T-1,\ldots, 1.
\end{aligned}
\end{equation}
\paragraph{Edit Intermediate-step Outputs of the Sampling Process.} The primary cause of inaccurately generated notes is the estimation error of the probability density of $\rmX_0$, which in turn affects the corresponding score function $\widehat{\vs}_t(\mX_t)$. The equivalence $\widehat{\vs}_t(\mX_t)=-\widehat{\vep}_\theta(\mX_t,t)/\sqrt{1-\bar{\alpha}_t}$ therefore inspires us to project $\widehat{\E}[\rmX_0|\mX_t]$ to the $\idK$-constrained domain $\sR^{L\times H}\backslash \sW_\idK$ by adjusting the value of $\widehat{\vep}_\theta(\mX_t,t)$. This adjustment is interpreted as an adjustment of the estimated score. Here $\sW_\idK$ is the set of matrices, connected to the set of positions (on the matrix) $w(\vl, \idK)$ by  
$$\sW_\idK = \left\{\mX\in\mathbb{R}^{L\times H}| ~\exists (l,h)\in w(\vl;\idK), \mX_{l,h}>1/2\right\}.$$

Specifically, at each sampling step $t$, we replace the guided noise prediction $\widehat{\vep}_\theta(\mX_t,t)$ with $\tilde{\vep}_\theta(\mX_t,t)$ such that 
\begin{equation}
\begin{aligned}
    \tilde{\vep}_\theta(\mX_t,t) = &\argmin_\vep \quad  \Vert \vep- \widehat{\vep}_\theta(\mX_t,t)\Vert\\
    \text{ s.t. } \quad & \left(\frac{\mX_t-\sqrt{1-\bar{\alpha}_t}\vep}{\sqrt{\bar{\alpha}_t}}\right)\in\sR^{L\times H}\backslash \sW^\prime_\idK.
\end{aligned}
\label{eq:optimization problem}
\end{equation}
The element-wise formulation of $\tilde{\vep}_\theta(\mX_t,t)$ is given as follows, with calculation details provided in Appendix \ref{appendix: prop2 proof}.
\begin{equation}
    \label{eq:corrected noise}
  \begin{aligned}
  \tilde{\vep}_{\theta,lh}&(\mX_t,t) = \vone\{(l,h)\not\in\omega(\vl;\idK)\}\cdot \widehat{\vep}_{\theta,lh}(\mX_t,t)\\
  +&\vone\{(l,h)\in\omega_\idK(\vl)\}\cdot\\
  &\max\left\{\widehat{\vep}_{\theta,lh}(\mX_t,t), \frac{1}{\sqrt{1-\bar{\alpha}_t}}\left(\emX_{t,lh}-\frac{\sqrt{\bar{\alpha}_t}}{2}\right)\right\}.
  \end{aligned}
\end{equation}
Plugging the adjusted noise prediction $\tilde{\vep}_\theta(\mX_t,t)$ into \eqref{eq: DDIM}, we derive the adjusted $\tilde{\rmX}_{t-1}$. The sampling process is therefore summarized as the following Algorithm \ref{alg:sample}.

\begin{algorithm}[htbp]
\SetAlgoLined
\KwIn{Input parameters: forward process variances $\beta_t$, $\bar{\alpha}_t=\prod_{s=1}^t\beta_t$, backward noise scale $\sigma_t$, key signature guidance $\idK$ }
\KwOut{generated piano roll $\tilde{\rmM}\in\{0,1\}^{L\times H}$}
$\rmX_T\sim\mathcal{N}(0,\mI)$;

\For{$t = T,T-1,\ldots,1$}{
    Compute guided noise prediction $\widehat{\vep}_\theta(\mX_t,t)$\;
    
    Perform noise correction:   derive $\tilde{\vep}_\theta(\mX_t,t)$ using \eqref{eq:corrected noise}\;

    Compute $\tilde{\rmX}_{t-1}$ by plugging the corrected noise $\tilde{\vep}_\theta(\mX_t,t)$ into \eqref{eq: DDIM}
}
Convert $\tilde{\rmX}_0$ into piano roll $\tilde{\rmM}$ 

\Return{$output$}\;
\caption{DDPM sampling with fine-grained harmonic control}
\label{alg:sample}
\end{algorithm}
Note that at the final step $t=0$, the noise correction directly projects $\widehat{\rmX}_0 $ to $\sR^{L\times H}\backslash\sW^\prime_\idK$, ensuring the probabilistic constraint \ref{eq:prob0}.

\paragraph{Theoretical Property of the Sampling Control.} A natural concern is that enforcing precise fine-grained control over generated samples may disrupt the learned local patterns. The following proposition \ref{prop:overall}, proved in \ref{appendix: proof SDE}, provides an upper bound that quantifies this potential effect and address the concern. 

\begin{proposition}
    \label{prop:overall}
    Under the SDE formulation in \eqref{eq:forward SDE} and \eqref{eq: reverse SDE}, given an early-stopping time $t_0$\footnote{We adopt the early-stopping time to avoid the blow-up of score function, which is standard in many literature \citep{song2020improved, nichol2021improved}}, if 
    \begin{equation}
        \mathbb{E}_{\rmX_t\sim p_t}[\Vert {\vep}^*(\rmX_t,t)-{\vep}_\theta(\rmX_t,t)\Vert^2]\leq \delta
    \end{equation} for all $t$, where ${\vep}^*(\rmX_t,t)$ is the optimal solution of the DDPM training objective (\ref{eq:loss function}), then we have
    \begin{equation*}
    \begin{aligned}
        &\text{KL}(\tilde{p}_{t_0}|p_{t_0})\leq \frac{\delta}{2}\int_{t_0}^T\frac{\beta(t)}{\sqrt{1-e^{-\int_{t_0}^t\beta(s)ds}}} dt,\quad\\
        &\text{KL}(\tilde{p}_{t_0}|\hat{p}_{t_0})\leq \frac{\delta}{2}\int_{t_0}^T\frac{\beta(t)}{\sqrt{1-e^{-\int_{t_0}^t\beta(s)ds}}} dt,
    \end{aligned}
    \end{equation*}
    where $p_{t_0}$ is the distribution of $\rmX_{t_0}$ in the forward process, $\hat{p}_{t_0}$ is the distribution of $\widehat{\rmX}_{t_0}$ generated by the diffusion sampling process without noise adjustment, and $\tilde{p}_{t_0}$ is the distribution of $\tilde{\rmX}_{t_0}$ generated by the fine-grained noise adjustment.
\end{proposition}

Proposition \ref{prop:overall} provides upper bounds for the distance between the controlled distribution and the uncontrolled distribution, as well as between the controlled distribution and the ground truth. We remark that, our method can shape the output towards a specific tonal quality. This can be for example using the Dorian scale as the key signature sequence $\idK$ to shape the generated music towards the Dorian mode (a tonal framework not present in the training data), where the generated distribution $\tilde{p}$ with fine-grained noise adjustment is fundamentally different from the ground truth distribution $p$. Nevertheless, Proposition \ref{prop:overall} guarantees a substantial overlap between the two distributions $\tilde{p}$ and $p$, demonstrating a well-balanced interplay between external control and the model's internal learning from the training data, e.g., melodic lines. This theoretical insight aligns with our empirical observations, which is presented in the ``Mode Change'' section of the demo page.

\section{Challenges for Uncontrolled Symbolic Music Generation Models}
\label{sec:challenges}

In the previous section \ref{sec:methodology}, we present our FGG method that guarantees the precision of generation. But why is it meaningful to provide such guarantee in the task of symbolic music generation? Why is it hard for models to self-ensure harmonic precision without having the hard sampling control? We use Section \ref{sec:challenges} to answer these questions. These discussions further motivate and justify the importance of the FGG method.

In the rest of this section, we focus our discussion to tonic-centric genres. Although not covering every aspect of music, it still spans a wide range of genres that are deeply embedded in everyday life, including tonic-centric New Age music, light classical music, and tonic-focused movie soundtracks. Such genres rely heavily on \textit{harmony}, i.e., the simultaneous sound of different notes that form a cohesive entity in the mind of the listener \citep{muller2015fundamentals}. 


Using the concept of temporary tonic key signatures we discussed in the previous section, we focus our discussion on the presence of out-of-key notes\footnote{For instance, a G$\natural$ note is considered as out-of-key in a G$\flat$ major context. Admittedly the inference of temporary tonic key is even more vague than chord recognition, due to the flexibility of harmony. However, in the following discussion, we assume that the temporary tonic key is specified.} in generated music. In the tonic-centric genres, out-of-key notes are uncommon, and produce noticeable dissonance, if not having a ``resolution''. We often notice that out-of-key notes are usually perceived merely as mistakes when appearing in generative model outputs, as demonstrated by some examples on our demo page. 

We aim to explain why the existence of out-of-key notes is an issue for diffusion-based symbolic music generation models in the tonic-centric genres. Specifically, we explain the following phenomenon: Suppose $\mathcal{G}$ is a diffusion model trained to generate tonic-centric genres. In the target data distribution, out-of-key notes appear at a small rate $P(O)\gtrsim 0$. These out-of-key notes are carefully managed (by expert composers) in the training set. However, when out-of-key notes appear in the generated samples of $\mathcal{G}$, they often lack an appropriate resolution and are more likely to be perceived negatively. Why does the model often fail to learn this nuance?

We provide an intuitive explanation using statistical reasoning. Consider a piano roll segment, represented as a random variable $\rmM$.  Suppose we are interested in whether this segment contains an out-of-key note (denoted as event $\{O\}$) and whether that note is eventually resolved within the segment (denoted as event 
$\{R,O\}$). In our training data, almost every out-of-key note is resolved, meaning the probability of unresolved out-of-key note is close to 0, i.e., $P(\bar{R},O)\approx 0$.

Now, we examine the probability 
 in the generated music. The key question then is whether the generative model also learns to keep $\widehat{P}(\bar{R},O)$ 
small. The following proposition \ref{prop:stat_error} leverages analysis of statistical errors to show that $\widehat{P}(\bar{R}, O)$ can decrease slowly as the dataset size $n$ increases.


\begin{proposition}
\label{prop:stat_error}
    Consider generating piano roll $\rmM$ from a continuous random variable $\rmX$, i.e.,  given $n$ i.i.d. data $\{\rmX^{i}\}_{i=1}^n\sim p_\rmX$, let $\{\rmM^{i}\}_{i=1}^n$ be given by  $\ermM^i_{lh} = \vone\{\ermX^i_{lh}\geq 1/2\}$. Denote the model for estimating the distribution of $\mathbf{X}$ as $\widehat{p}_\rmX$. We have  $\exists~C>0$ such that $\forall n$,
\begin{equation}
\label{eq:decay}\inf_{\widehat{p}_\rmX}\sup_{p_\rmX\in\mathcal{P}_\delta}\E_{\{\rmM^i\}_{i=1}^n}\widehat{P}(\bar{R},O)\geq C\cdot
 n^{-\frac{1}{LH+2}}-P(\bar{R},O),
\end{equation}
where $\widehat{P}$ is the probability associated with the generated data $\estpx$.
\end{proposition}

The proof of proposition \ref{prop:stat_error} is provided in appendix \ref{appendix: prop1 proof}.
The term $\sup_{p_\rmX\in\mathcal{P}_\delta}$ is the supremum taken over the  search space of the continuous generative model\footnote{The exact formulation of $P_\delta$ is given in appendix \ref{appendix: prop1 proof}. While real life distribution classes associated with generative models are more complicated and difficult to analyze, $P_\delta$ essentially captures their characteristics, and is therefore comparable to them. This type of simplification while maintaining core characteristics appears to be standard in works that provide theoretical insights \cite{fu2024unveil}.}, and $\inf_{\widehat{p}_\rmX}$ denotes the best possible realization of the model. The minimax formulation is standard in 
works that discuss statistical convergence of generative models \cite{fu2024unveil}.

The theoretical insights presented in proposition \ref{prop:stat_error} demonstrate that the occurrence of unsolved out-of-key note is often unavoidable, and the decay rate of this error probability with respect to training set size $n$ is slow $O(n^{-1/(LH)})$. Thus, relying on the model itself for precision is challenging for existing models, given the inherent scarcity of high-quality data and the slow decay rate of errors. There are two implications following this line: First, it would be immensely valuable to develop a model that enjoys the ability to implicitly learn contextually appropriate out-of-key notes (nevertheless, currently in our work we did not take this path). Second, with the fact that symbolic music generation requires an exceptional level of precision, it is worthwhile to develop methods that enable the model to function as a well-controlled collaborative tool to aid human composers. 


\section{Experiments}
\label{sec:experiments}
In this section, we present  experiments to demonstrate the effectiveness of our fine-grained guidance approach. We additionally create a demopage\footnote{ See \href{https://huajianduzhuo-code.github.io/FGG-diffusion-music/}{https://huajianduzhuo-code.github.io/FGG-diffusion-music/}. We note that slow performance may result from Huggingface resource limitations and network latency.} for demonstration, which allows for fast and stable interactive music creation with user-specified input guidance\footnote{The format of user-specified input guidance is limited within the constrained piano roll format, as is demonstrated in the paper.}, and even for generating music based on tonal frameworks absent from the training set.

\subsection{Numerical Experiments}
\label{subsec:experiment}
We present numerical experiments on accompaniment generation given both melody and chord generation, or symbolic music generation given only chord conditions. We focus on the former one as it provides a more effective basis for comparison. Due to page limits, we put the results and more detailed explanation of the latter one in Appendix~\ref{appendix:music generation task}. For the accompaniment generation task, we compare with two state-of-the-art baselines: 1) WholeSongGen \citep{wang2024whole} and 2) GETMusic \citep{lv2023getmusic}.

\subsubsection{Data Representation and Model Architecture}
\label{sec:model architecture}
The generation target $\mX$ is represented by a piano-roll matrix of shape $2\times L\times 128$ under the resolution of a 16th note, where $L$ represents the total length of the music piece, and the two channels represent note onset and sustain, respectively. In our experiments, we set $L=64$, corresponding to a 4-measure piece under time signature $4/4$. Longer pieces can be generated autoregressively using the inpainting method. The backbone of our model is a 2D UNet with spatial attention.

The condition matrix $\mMc$ is also represented by a piano roll matrix of shape $2\times L\times 128$, with the same resolution and length as that of the generation target $\mX$. For the accompaniment generation experiments, we provide melody as an additional condition. Detailed construction of the condition matrices are provided in Appendix \ref{appendix:data representation details}.

\subsubsection{Dataset}
We use the POP909 dataset \cite{wang2020pop909} for training and evaluation. This dataset consists of 909 MIDI pieces of pop songs, each containing lead melodies, chord progression, and piano accompaniment tracks. We exclude 29 pieces that are in triple meter. 90\% of the data are used to train our model, and the remaining 10\% are used for evaluation. In the training process, we split all the midi pieces into 4-measure non-overlapping segments (corresponding to $L=64$ under the resolution of a 16th note), which in total generates 15761 segments in the entire training set. Training and sampling details are provided in Appendix~\ref{appendix:training and sampling details}.

\subsubsection{Task and Baseline Models}
We consider accompaniment generation task based on melody and chord progression. We compare the performance of our model with two baseline models: 1) {WholeSongGen} \citep{wang2024whole} and 2) {GETMusic} \citep{lv2023getmusic}. {WholeSongGen} is a hierarchical music generation framework that leverages cascaded diffusion models to generate full-length pop songs. It introduces a four-level computational music language, with the last level being accompaniment. The model for the last level can be directly used to generate accompaniment given music phrases, lead melody, and chord progression information. {GETMusic} is a versatile music generation framework that leverages a discrete diffusion model to generate tracks based on flexible source-target combinations. The model can also be directly applied to generate piano accompaniment conditioning on melody and chord. Since these baseline models do not support rhythm control, to ensure comparability, we will use the $\mMc(\mathcal{C})$ without rhythm condition in our model.
\subsubsection{Evaluation}
We generate accompaniments for the 88 MIDI pieces in our evaluation dataset.\footnote{The WholeSongGen model from \citet{wang2024whole} is also trained on the POP909 dataset. Our evaluation set is a subset of their test set so there is no in-sample evaluation issue on their model.} We introduce the following objective metrics to evaluate the generation quality of different methods:

(1) \textit{Percentage of Out-of-Key Notes} First, for each method, we present the frequency of out-of-key notes by computing the percentage of steps in the generated sequences containing at least one out-of-key note, where each step corresponds to a 16th note. The results, presented in Table~\ref{tab:evaluation-table}, indicate that frequency of out-of-key notes in the baselines is roughly 2\%-4\%, equating to about 1–3 occurrences in a 4-measure piece. In contrast, our sampling control method effectively eliminates such dissonant notes in the generated samples.

(2) \textit{Direct Chord Accuracy} and \textit{Chord Progression Similarity} We evaluate harmonic consistency by comparing the chord progressions of the generated and ground truth accompaniments. Chords are extracted using the rule-based recognition method from \citet{dai2020automatic}. Direct chord accuracy is computed as the percentage of beats where the recognized chord of the generated output exactly matches that of the ground truth. However, since not all mismatches reflect equal harmonic deviation—for instance, C major is harmonically close to Cmaj7 but far from B major—direct accuracy may fail to reflect the nuanced similarity between chords.

To address this, we further assess chord progression similarity. We divide each accompaniment into non-overlapping 2-measure segments and encode them into a 256-dimensional latent space using a pre-trained disentangled VAE \cite{wang2020learning}. Cosine similarity is then computed between corresponding segments from the generated and ground truth progressions. Table~\ref{tab:evaluation-table} reports the average direct accuracy and average latent similarity, along with their 95\% confidence intervals. The results demonstrate that our method significantly outperforms all baselines in chord accuracy.

(3) \textit{Intersection over Union (IoU) Metrics.}
We evaluate the similarity between the generated and ground truth accompaniments using two IoU-based metrics: \textit{IoU of Chords} and \textit{IoU of Piano Roll}. For \textit{IoU of Chords}, we first apply the chord recognition method from \citet{dai2020automatic} to both the generated and ground truth accompaniments. Each chord is then represented as a 12-dimensional binary vector indicating the presence of pitch classes (C through B). We compute the IoU between the generated and ground truth pitch class sets at every 16th-note time step and report the average IoU across all time steps.

For \textit{IoU of Piano Roll}, we represent each accompaniment as a binary piano roll. The IoU is computed at each 16th-note time step by comparing the sets of active pitches in the generated and ground truth piano rolls. We then report the average IoU across all time steps.\footnote{While exact agreement with the ground truth is not necessarily optimal—since a given melody may admit multiple valid accompaniments—the IoU still serves as a useful indicator of musical quality. A high-quality accompaniment is expected to align reasonably well with the expert-written ground truth, and thus should not deviate substantially.} The results, presented in Table~\ref{tab:evaluation-table}, show that our method consistently achieves higher IoU scores than the baselines, indicating closer alignment to the ground truth at both the harmonic and note level.

\begin{table*}[ht]
\begin{center}
\adjustbox{max width=\linewidth}{
\renewcommand{\arraystretch}{1.2}
\setlength{\tabcolsep}{10pt}
\small
\begin{tabular}{l|clllllll}
\hline
\rowcolor[HTML]{D9EAD3}
\multicolumn{1}{c|}{\bf Methods} &\multicolumn{1}{c}{\bf \% Out-of-Key Notes $\downarrow$} & \multicolumn{1}{c}{\bf Direct Chord Accuracy $\uparrow$} &\multicolumn{1}{c}{\bf Chord Similarity $\uparrow$} & \multicolumn{1}{c}{\bf IoU (Chord) $\uparrow$} & \multicolumn{1}{c}{\bf IoU (Piano Roll) $\uparrow$} \\
\hline
\rowcolor[HTML]{FFFFFF}
\textbf{FGG (Ours)} &0.0\% & \bm{$0.485\pm 0.006$} & $\bm{0.767\pm 0.007}$ & \bm{$0.769\pm 0.003$} & \bm{$0.281\pm 0.005$}\\
\rowcolor[HTML]{F4F4F4}
WholeSongGen  &2.1\% & $0.314\pm 0.006$ & $0.611\pm 0.010$ & $0.618\pm 0.004$ & $0.107\pm 0.003$\\
\rowcolor[HTML]{FFFFFF}
GETMusic      &3.5\% & $0.153\pm 0.007$ & $0.394\pm 0.012$ & $0.412\pm 0.007$ & $0.048\pm 0.003$\\
\hline
\end{tabular}
}
\end{center}
\caption{Evaluation of the similarity with ground truth for all methods.}
\label{tab:evaluation-table}
\end{table*}

\textit{(4) Subjective Evaluation} 

\begin{figure*}[t]
    \centering
    \includegraphics[width=0.9\linewidth]{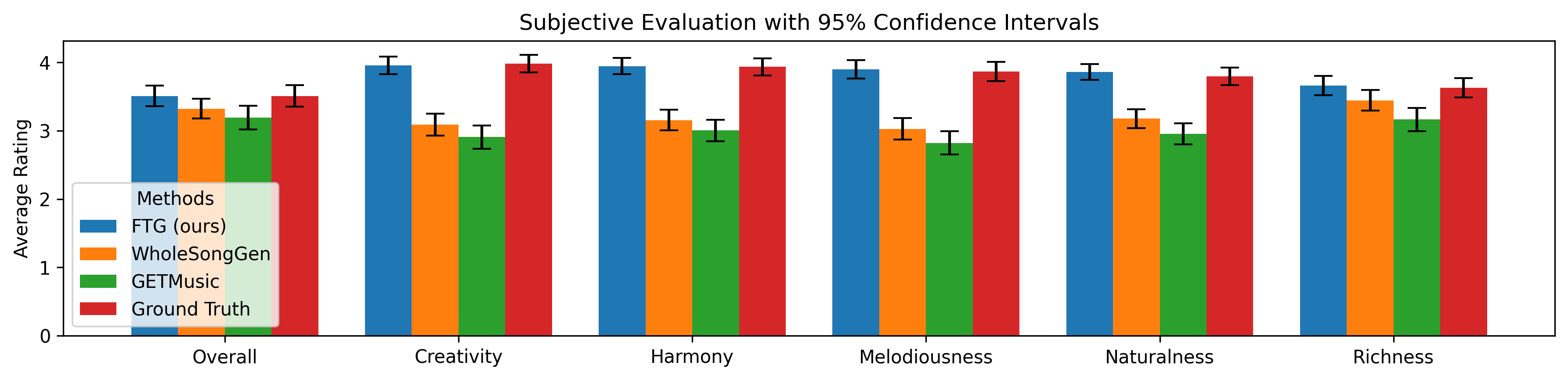}
    \caption{Subjective evaluation results on music quality.}
    \label{fig:subjective}
\end{figure*}

To compare performance of our FGG method against the baselines (ground truth, WholeSongGen, and GETMusic), we prepared 6 sets of generated samples, with each set containing the melody paired with accompaniments generated by FGG, WholeSongGen, and GETMusic, along with the ground truth accompaniment. This yields a total of $6\times 4 = 24$ samples. The samples are presented in a randomized order, and their sources are not disclosed to participants. Experienced listeners assess the quality of samples
in 5 dimensions: creativity, harmony (whether the accompaniment is in harmony with the melody),
melodiousness, naturalness and richness, together with an overall assessment. The results are shown in Figure \ref{fig:subjective}. The bar height shows the mean rating, and the error bar shows the 95\% confidence interval. FGG
consistently outperforms the baselines in all dimensions.
For details of our survey, please see Appendix \ref{appendix:subjective}.

\subsubsection{Ablation Study}
\label{sec:ablation study}
In this section, we conduct ablation studies to better illustrate the effectiveness of our FGG method. We aim to demonstrate the effectiveness of both the fine-grained training condition (training control) and the sampling control. We also compare with simple rule-based post-sample editing\footnote{Specifically, we compare with two rule-based post-sample editing methods: 1) Rounding wrong notes up/down and 2) Remove wrong notes.}. The former leverages the structured gradual denoising process of diffusion models, ensuring a theoretical guarantee of preserving the distributional properties of the original learned distribution. In contrast, the latter employs a brute-force editing approach that disrupts the generated samples, affecting local melodic lines and rhythmic patterns. The numerical results further validate this analysis.

Moreover, we compare with a so-called ``Inpainting'' method, which treats the pixels where there is not supposed to be a note as 0 and inpaints the remaining pixels. This information is included by adding a mask channel in the training process. We still allow for the fine-grained conditioning in training.

Specifically, we include the following variants in our ablation study:
\begin{itemize}
    \item \textbf{Training and Sampling Control}: our full method, which applies fine-grained conditioning during both training and sampling.
    \item \textbf{Inpainting}: out-of-key pixels are treated as 0, and the remaining positions are inpainted based on fixed context.
    \item \textbf{Training control + Round Notes Up/Down After Sampling}: training control is applied, and out-of-key notes are corrected post-sampling by rounding to the nearest in-key pitch.
    \item \textbf{Training control + Remove Wrong Notes After Sampling}: training control is applied, and out-of-key notes are corrected post-sampling by removing them.
    \item \textbf{Training Control Only}: only training control is used; no sampling-time controls are enforced.
    \item \textbf{No Control}: neither training control nor sampling control are applied.
\end{itemize}

We assess overall model performance using the same quantitative metrics as in the previous section. The results are shown in Table~\ref{tab:comparison-table}. In general, fine-grained conditioning (i.e., training control) leads to substantial improvements across all evaluation metrics, and adding sampling control further enhances performance. While rule-based post-sampling editing (e.g., removing or rounding out-of-key notes) yields moderate gains, it is consistently outperformed by our fine-grained sampling control method. Our approach fully leverages the structured, gradual denoising process of diffusion models, allowing the model to iteratively correct or replace errors while preserving the coherence of the original learned distribution.

Moreover, our method outperforms the inpainting baseline across all evaluation metrics. Unlike our approach, the inpainting method introduces additional architectural complexity by requiring the model to handle an auxiliary mask channel that indicates which pixels to regenerate. This not only increases implementation overhead but also adds computational burden during both training and inference.

\begin{table*}[!htbp]
\begin{center}
\adjustbox{max width=\linewidth}{
\renewcommand{\arraystretch}{1.2}
\setlength{\tabcolsep}{10pt}
\small
\begin{tabular}{c|ccccccc}
\hline
\rowcolor[HTML]{D9EAD3}  
\multicolumn{1}{c|}{\bf Methods} & \multicolumn{1}{c}{\bf \% Out-of-Key} & \multicolumn{1}{c}{\bf \textcolor{black}{Direct Chord} } & \multicolumn{1}{c}{\bf Chord} & \multicolumn{1}{c}{\bf \textcolor{black}{IoU} } & \multicolumn{1}{c}{\bf \textcolor{black}{IoU} }\\
\rowcolor[HTML]{D9EAD3}
 & \bf Notes & \bf Accuracy & \bf Similarity & \bf Chord & \bf Piano Roll \\
\hline

\rowcolor[HTML]{FFFFFF}
\textbf{Training and } & 0.0\% & \textcolor{black}{\bm{$0.485$}} & \bm{$0.767$} & \textcolor{black}{\bm{$0.769$}} & \textcolor{black}{\bm{$0.281$}} \\
\rowcolor[HTML]{FFFFFF}
\textbf{Sampling Control} & & \textcolor{black}{\bm{$\pm 0.006$}} & \bm{$\pm 0.007$} & \textcolor{black}{\bm{$\pm 0.003$}} & \textcolor{black}{\bm{$\pm 0.005$}} \\

\rowcolor[HTML]{F4F4F4}
Inpainting & 0.0\% & $0.458$ & $0.710$ & \textcolor{black}{$0.743$} & \textcolor{black}{$0.271$} \\
\rowcolor[HTML]{F4F4F4}
 & & \textcolor{black}{$\pm 0.006$} & $\pm 0.008$ & \textcolor{black}{$\pm 0.003$} & \textcolor{black}{$\pm 0.005$} \\

\rowcolor[HTML]{FFFFFF}
\textcolor{black}{Training Control} & \textcolor{black}{0.0\%} & \textcolor{black}{$0.472$} & \textcolor{black}{$0.756$} & \textcolor{black}{$0.763$} & \textcolor{black}{$0.272$} \\
\rowcolor[HTML]{FFFFFF}
 \textcolor{black}{Round Notes Up/Down After Sampling} & & \textcolor{black}{$\pm 0.006$} & \textcolor{black}{$\pm 0.007$} & \textcolor{black}{$\pm 0.003$} & \textcolor{black}{$\pm 0.005$} \\

\rowcolor[HTML]{F4F4F4}
Training Control & 0.0\% & \textcolor{black}{$0.482$} & $0.763$ & \textcolor{black}{$0.767$} & \textcolor{black}{$0.277$} \\
\rowcolor[HTML]{F4F4F4}
Remove Wrong Notes After Sampling & & \textcolor{black}{$\pm 0.006$} & $\pm 0.007$ & \textcolor{black}{$\pm 0.003$} & \textcolor{black}{$\pm 0.005$} \\

\rowcolor[HTML]{FFFFFF}
Only & 3.7\% & \textcolor{black}{$0.465$} & $0.748$ & \textcolor{black}{$0.757$} & \textcolor{black}{$0.270$} \\
\rowcolor[HTML]{FFFFFF}
Training Control & & \textcolor{black}{$\pm 0.006$} & $\pm 0.007$ & \textcolor{black}{$\pm 0.003$} & \textcolor{black}{$\pm 0.005$} \\

\rowcolor[HTML]{F4F4F4}
No Control & 10.1\% & \textcolor{black}{$0.112$} & $0.378$ & \textcolor{black}{$0.378$} & \textcolor{black}{$0.072$} \\
\rowcolor[HTML]{F4F4F4}
 & & \textcolor{black}{$\pm 0.004$} & $\pm 0.007$ & \textcolor{black}{$\pm 0.004$} & \textcolor{black}{$\pm 0.002$} \\

\hline
\end{tabular}
}
\end{center}
\caption{Ablation study.}
\label{tab:comparison-table}
\end{table*}

\subsection{Empirical Observations}
Notably, harmonic control not only helps the model eliminate incorrect notes, but also guides it to replace them with correct ones. Such representative examples are presented in Appendix \ref{appendix:empirical}. Moreover, samples generated from ablation conditions are available in Section 3 of our demo page\footnote{The demo page is available at \href{https://huajianduzhuo-code.github.io/FGG-diffusion-music/}{https://huajianduzhuo-code.github.io/FGG-diffusion-music/}}. Across all ablations, we observed occasional occurrences of excessively high-pitched notes and overly dense note clusters.

\section{Limitations and Future Work}
\label{sec:limitations}

While our method achieves strong performance across multiple metrics, several limitations remain. First, we adopt a 16th-note quantization scheme following \citet{wang2024whole}, which simplifies temporal representation but restricts rhythmic flexibility and precludes training on data without explicit beat annotations. A promising future direction is to integrate our pitch-class-based control mechanism with approaches such as \citet{huang2024symbolic}, which introduce a dynamic dimension and utilize finer 10ms time quantization to better capture expressive timing variations. Second, our method supports pitch class and rhythmic control in the piano roll representation, but does not accommodate more abstract forms or probabilistic control. Finally, we note that evaluation remains a broader challenge across the field of symbolic music generation. Since musical quality evaluation is inherently detailed and partly subjective, objective evaluation metrics such as rule-based and structural evaluation methods used in this work have inherent limitations in reflecting perceptual or creative aspects of music. This is a key reason why many recent studies supplement objective evaluation with subjective human listening evaluations. A valuable future direction is to develop improved automatic evaluation metrics that more faithfully align with human judgments of musicality and creativity.

\section{Conclusion}
In this work, we apply fine-grained textural guidance (FGG) on symbolic music generation models. We provide theoretical analysis and empirical evidence to highlight the need for fine-grained and precise control over the model output. We also provide theoretical analysis to quantify and upper bound the potential effect of fine-grained control on learned local patterns, and provide samples and numerical results for demonstrating the effectiveness of our approach. For the impact of our method, we note that the FGG method can be integrated with other diffusion-based symbolic music generation methods. With a moderate trade-off of flexibility, the FGG method prioritizes real-time generation stability and enables efficient generation with precise control.

 \section*{Acknowledgements}
The authors gratefully acknowledge Jinghai He, Ang Lv, Yifu Tang, Gus Xia, Yaodong Yu, Yufeng Zheng, anonymous reviewers, area chairs, and the anonymous evaluators of this work's demos. 


\section*{Impact Statement}
This paper presents work whose goal is to advance the field of 
Machine Learning. There are a range of potential societal consequences 
of our work, none which we feel must be specifically highlighted here.


\bibliography{icml2025}
\bibliographystyle{icml2025}

\newpage
\appendix
\onecolumn

\section{Proof of propositions and calculation details}

\subsection{Calculation details in \ref{subsec:fine-grained sampling}}
\label{appendix: prop2 proof}

Our goal is to find the optimal solution of problem (\ref{eq:optimization problem}). Since the constraint is an element-wise constraint on a linear function of $\vep$ and the objective is separable, we can find the optimal solution by element-wise optimization. Consider the $(l,h)$-element of $\vep$.

First, if $(l,h)\notin w(\vl;\idK)$, there is no constraint on $\vep_{lh}$. Therefore, the optimal solution of $\vep_{lh}$ is $\widehat{\vep}_{\theta,lh}(\mX_t,t)$.

If $(l,h)\in w(\vl;\idK)$, the constraint on $\vep_{lh}$ is
\begin{equation*}
    X_{t,lh}-\frac{\sqrt{1-\bar{\alpha}_t}\vep_{lh}}{\sqrt{\bar{\alpha}_t}}\leq \frac{1}{2},
\end{equation*}
which is equivalent to
\begin{equation*}
    \vep_{lh}\geq \frac{1}{\sqrt{1-\bar{\alpha}_t}}\left(\emX_{t,lh}-\frac{\sqrt{\bar{\alpha}_t}}{2}\right).
\end{equation*}

The objective is to minimize $\Vert \vep_{lh}- \widehat{\vep}_{\theta,lh}(\mX_t,t)\Vert$. Therefore, the optimal solution of $\vep_{lh}$ is $$\vep_{lh}=\max\left\{\widehat{\vep}_{\theta,lh}(\mX_t,t|\idC,\idR), \frac{1}{\sqrt{1-\bar{\alpha}_t}}\left(\emX_{t,lh}-\frac{\sqrt{\bar{\alpha}_t}}{2}\right)\right\}.$$

\subsection{Proof of Proposition \ref{prop:overall}}
\label{appendix: proof SDE}
\begin{proof}
Recall that According to \citet{song2020score}, the DDPM forward process $\rmX_t=\sqrt{\bar{\alpha}_t}\rmX_0+\sqrt{1-\bar{\alpha}_t}{\vep}$ can be regarded as a discretization of the following SDE:
\begin{equation*}
    d\rmX_t = -\frac{1}{2}\beta(t)\rmX_t dt+\sqrt{\beta(t)} d\rmW_t,
\end{equation*}

and the corresponding denoising process takes the form of a solution to the following stochastic differential equation (SDE):
\begin{equation*}
    d\rmX_t = -\left[\frac{1}{2}\beta(t)\rmX_t+\beta(t)\nabla_{\rmX_t}\log p_t(\rmX_t)\right]dt+\sqrt{\beta(t)}d\bar{\rmW}_t,
\end{equation*}
where $\beta(t/T)=T\beta_t$ as $T$ goes to infinity, $\bar{\rmW}_t$ is the reverse time standard Wiener process, and $\bar{\alpha}_t$ term should be replaced by its continuous version $e^{-\int_0^t\beta(s)ds}$ (or $e^{-\int_{t_0}^t\beta(s)ds}$ when early-stopping time $t_0$ is adopted). The score function $\nabla_{\rmX_t}\log p_t(\rmX_t)$ can be approximated by $-\vep_\theta(\rmX_t,t)/\sqrt{1-e^{-\int_0^t\beta(s)ds}}$.

Under the SDE formulation, the denoising process can take the form of a solution to stochastic differential equation (SDE):

\begin{equation}
\label{eq: reverse SDE_appendix}
    d\rmX_t = -\left[\frac{1}{2}\beta(t)\rmX_t+\beta(t)\nabla_{\rmX_t}\log p_t(\rmX_t)\right]dt+\sqrt{\beta(t)}d\bar{\rmW}_t,
\end{equation} 
where $\beta(t/T)=T\beta_t$, $\bar{\rmW}_t$ is the reverse time standard Wiener process. According to \citet{song2020score}, as $T\rightarrow\infty$, the solution to the SDE converges to the real data distribution $p_0$.

In the diffusion model, $\nabla_{\rmX_t}\log p_t(\rmX_t)$ is approximated by $-\vep_\theta(\rmX_t,t)/\sqrt{1-e^{-\int_{t_0}^t\beta(s)ds}}$. Therefore, the approximated reverse-SDE sampling process without harmonic guidance is 
\begin{equation}
\label{eq: reverse SDE_approx}
    d\hat{\rmX}_t = -\left[\frac{1}{2}\beta(t)\rm\hat{\rmX}_t-\beta(t)\frac{\vep_\theta(\hat{\rmX}_t,t)}{\sqrt{1-e^{-\int_{t_0}^t\beta(s)ds}}}\right]dt+\sqrt{\beta(t)}d\bar{\rmW}_t.
\end{equation}

Similarly, the sampling process with fine-grained harmonic guidance is
\begin{equation}
    d\tilde{\rmX}_t = -\left[\frac{1}{2}\beta(t)\rm\tilde{\rmX}_t-\beta(t)\frac{\tilde{\vep}_\theta(\tilde{\rmX}_t,t)}{\sqrt{1-e^{-\int_{t_0}^t\beta(s)ds}}}\right]dt+\sqrt{\beta(t)}d\bar{\rmW}_t,
\end{equation}
where $\tilde{\vep}_\theta$ is defined as \eqref{eq:optimization problem}
and \eqref{eq:corrected noise}.

For simplicity, we denote the drift terms as follows:
\begin{align*}
    f(\rmX_t,t) &= -\left[\frac{1}{2}\beta(t)\rmX_t+\beta(t)\nabla_{\rmX_t}\log p_t(\rmX_t)\right]\\
    \hat{f}(\hat{\rmX}_t,t) &= -\left[\frac{1}{2}\beta(t)\rm\hat{\rmX}_t-\beta(t)\frac{\vep_\theta(\hat{\rmX}_t,t)}{\sqrt{1-e^{-\int_{t_0}^t\beta(s)ds}}}\right],\\
    \tilde{f}(\tilde{\rmX}_t,t) &= -\left[\frac{1}{2}\beta(t)\rm\tilde{\rmX}_t-\beta(t)\frac{\tilde{\vep}_\theta(\tilde{\rmX}_t,t)}{\sqrt{1-e^{-\int_{t_0}^t\beta(s)ds}}}\right].
\end{align*}

Since $$\mathbb{E}_{\rmX_t\sim p_t}[\Vert {\vep}^*(\rmX_t,t)-{\vep}_\theta(\rmX_t,t)\Vert^2]\leq \delta,$$ and $${\vep}^*(\rmX_t,t)=-\sqrt{1-e^{-\int_{t_0}^t\beta(s)ds}}\nabla_{\rmX_t}\log p_t(\rmX_t),$$ we have
\begin{align*}
    \mathbb{E}_{\rmX\sim p_t}[\Vert f(\rmX,t)-\hat{f}(\rmX,t)\Vert]\leq \frac{\beta(t)}{\sqrt{1-e^{-\int_{t_0}^t\beta(s)ds}}}\delta.
\end{align*}

Now we consider $\tilde{\vep}_\theta(\tilde{\rmX}_t,t)$, which is the solution of the optimization problem (\ref{eq:optimization problem}). In the continuous SDE case, the corresponding optimization problem becomes

\begin{equation}
\begin{aligned}
    \min_\vep \quad & \Vert \vep- \widehat{\vep}_\theta(\mX_t,t|\idC,\idR)\Vert\\
    \text{ s.t. } \quad & \left(\frac{\mX_t-\sqrt{1-e^{-\int_{t_0}^t\beta(s)ds}}\vep}{e^{-\frac{1}{2}\int_{t_0}^t\beta(s)ds}}\right)\in\sR^{L\times H}\backslash \sW^\prime_\idK.
\end{aligned}
\label{eq:optimization problem continuous}
\end{equation}

According to Proposition 1 of \citet{chung2022diffusion}, the posterior mean of $\rmX_0$ conditioning on $\rmX_t$ is
\begin{equation*}
\begin{aligned}
    \mathbb{E}[\rmX_0|\rmX_t] &= \frac{1}{e^{-\frac{1}{2}\int_{t_0}^t\beta(s)ds}}\left(\rmX_t+(1-e^{-\frac{1}{2}\int_{t_0}^t\beta(s)ds})\nabla_{\rmX_t}\log p_t(\rmX_t)\right)\\
    &= \frac{1}{e^{-\frac{1}{2}\int_{t_0}^t\beta(s)ds}}\left(\rmX_t-\sqrt{1-e^{-\int_{t_0}^t\beta(s)ds}}\vep^*(\rmX_t,t) \right).
\end{aligned}
\end{equation*}
Since the domain of $\rmX_0$ is $R^{L\times H}\backslash \sW^\prime_\idK$, which is a convex set, we know that the posterior mean $\mathbb{E}[\rmX_0|\rmX_t]$ naturally belongs to its domain. Therefore, $\vep^*(\rmX_t,t)$ is feasible to the problem (\ref{eq:optimization problem continuous}). Since the optimal solution of the problem is $\tilde{\vep}_\theta(\rmX_t,t)$, we have
\begin{equation*}
    \Vert \tilde{\vep}_\theta(\rmX_t,t)- \vep_\theta(\rmX_t,t)\Vert \leq \Vert \vep^*(\rmX_t,t)- \vep_\theta(\rmX_t,t)\Vert
\end{equation*}
for all $\rmX_t$ and $t$. This further leads to the result that
\begin{equation}
\label{eq:bound1}
    \mathbb{E}_{\rmX\sim p_t}[\Vert \tilde{f}(\rmX,t)-\hat{f}(\rmX,t)\Vert]\leq \frac{\beta(t)}{\sqrt{1-e^{-\int_{t_0}^t\beta(s)ds}}}\delta.
\end{equation}

Moreover, since $\tilde{\vep}_\theta(\rmX_t,t)$ is essentially the projection of $\vep_\theta(\rmX_t,t)$ onto the convex set defined by the constraints in (\ref{eq:optimization problem continuous}), and $\vep^*(\rmX_t,t)$ also belongs to the set, we know that the inner product of $\vep^*(\rmX_t,t)-\tilde{\vep}_\theta(\rmX_t,t)$ and $\vep_\theta(\rmX_t,t)-\tilde{\vep}_\theta(\rmX_t,t)$ is negative, which further leads to the result that
\begin{equation}
    \Vert \tilde{\vep}_\theta(\rmX_t,t)- \vep^*(\rmX_t,t)\Vert \leq \Vert \vep^*(\rmX_t,t)- \vep_\theta(\rmX_t,t)\Vert,
\end{equation}
which further implies
\begin{equation}
\label{eq:bound2}
    \mathbb{E}_{\rmX\sim p_t}[\Vert \tilde{f}(\rmX,t)-f(\rmX,t)\Vert]\leq \frac{\beta(t)}{\sqrt{1-e^{-\int_{t_0}^t\beta(s)ds}}}\delta.
\end{equation}

The following Girsanov's Theorem \cite{karatzas1991brownian} will be used (together with \eqref{eq:bound1} and \eqref{eq:bound2}) to prove the upper bounds for the KL-divergences in our Proposition~\ref{prop:overall}:

\begin{proposition}
\label{prop:girsanov}
Let $p_0$ be any probability distribution, and let $Z = (Z_t)_{t \in [0, T]}$, $Z' = (Z'_t)_{t \in [0, T]}$ be two different processes satisfying
\begin{align*}
     dZ_t &= b(Z_t, t) dt + \sigma(t) dB_t, \quad Z_0 \sim p_0,\\
     dZ'_t &= b'(Z'_t, t) dt + \sigma(t) dB_t, \quad Z'_0 \sim p_0.
\end{align*}

We define the distributions of $Z_t$ and $Z'_t$ as $p_t$ and $p'_t$, and the path measures of $Z$ and $Z'$ as $\mathbb{P}$ and $\mathbb{P}'$ respectively.

Suppose the following Novikov's condition:
\begin{equation}
\label{eq:novikov's condition}
    \mathbb{E}_{\mathbb{P}} \left[ \exp \left( \int_0^T \frac{1}{2} \int_x \sigma^{-2}(t) \| (b - b')(x, t) \|^2 dx dt \right) \right] < \infty.
\end{equation}
Then, the Radon-Nikodym derivative of $\mathbb{P}$ with respect to $\mathbb{P}'$ is
\begin{align*}
    \frac{d\mathbb{P}}{d\mathbb{P}'}(Z) = \exp \left\{ -\frac{1}{2} \int_0^T \sigma(t)^{-2} \| (b - b')(Z_t, t) \|^2 dt - \int_0^T \sigma(t)^{-1} (b - b')(Z_t, t) dB_t \right\},
\end{align*}
and therefore we have that
\begin{align*}
    \text{KL}(p_T \| p'_T) \leq \text{KL}(\mathbb{P} \| \mathbb{P}') = \int_0^T \frac{1}{2} \int_x p_t(x) \sigma(t)^{-2} \| (b - b')(x, t) \|^2 dx dt.
\end{align*}
Moreover, \citet{chen2022sampling} showed that if $\int_x p_t(x) \sigma^{-2}(t) \| (b - b')(x, t) \|^2 dx \leq C$ holds for some constant $C$ over all $t$, we have that
\[
    \text{KL}(p_T \| p'_T) \leq \int_0^T \frac{1}{2} \int_x p_t(x) \sigma(t)^{-2} \| (b - b')(x, t) \|^2 dx dt,
\]
even if the Novikov's condition \eqref{eq:novikov's condition} is not satisfied.
\end{proposition}.

According to \eqref{eq:bound1} and \eqref{eq:bound2}, we have
\begin{align}
    \int_x p_t(x)\beta(t)^{-1} \Vert \tilde{f}(\rmX,t)-\hat{f}(\rmX,t)\Vert dx&\leq \frac{\beta(t)}{\sqrt{1-e^{-\int_{t_0}^t\beta(s)ds}}}\delta\leq \sup_{t\in[t_0,T]}\frac{\beta(t)}{\sqrt{1-e^{-\int_{t_0}^t\beta(s)ds}}}\delta,\\
    \int_x p_t(x)\beta(t)^{-1} \Vert \tilde{f}(\rmX,t)-f(\rmX,t)\Vert dx&\leq \frac{\beta(t)}{\sqrt{1-e^{-\int_{t_0}^t\beta(s)ds}}}\delta\leq \sup_{t\in[t_0,T]}\frac{\beta(t)}{\sqrt{1-e^{-\int_{t_0}^t\beta(s)ds}}}\delta.
\end{align}
Therefore, we can apply Proposition~\ref{prop:girsanov} to obtain upper bounds for the KL-divergences, which leads to

\begin{equation}
\begin{aligned}
    \text{KL}(\tilde{p}_{t_0}|\hat{p}_{t_0})&\leq \int_{t_0}^T\frac{1}{2} \int_x p_t(x)\beta(t)^{-1} \Vert \tilde{f}(\rmX,t)-\hat{f}(\rmX,t)\Vert dx\\
    &\leq \delta \int_{t_0}^T\frac{1}{2}\frac{\beta(t)}{\sqrt{1-e^{-\int_{t_0}^t\beta(s)ds}}} dt
\end{aligned}
\end{equation}
and
\begin{equation}
\begin{aligned}
    \text{KL}(\tilde{p}_{t_0}|p_{t_0})&\leq \int_{t_0}^T\frac{1}{2} \int_x p_t(x)\beta(t)^{-1} \Vert \tilde{f}(\rmX,t)-f(\rmX,t)\Vert dx\\
    &\leq \delta \int_{t_0}^T\frac{1}{2}\frac{\beta(t)}{\sqrt{1-e^{-\int_{t_0}^t\beta(s)ds}}} dt.
\end{aligned}
\end{equation}    
\end{proof}

\begin{remark}
\label{remark:early stop}
    Under the SDE formulation, the forward process terminates at a sufficiently large time $T$. Also, since the score functions blow up at $t\approx 0$, an early-stopping time $t_0$ is commonly adopted to avoid such issue \cite{song2020improved, nichol2021improved}. When $t_0$ is sufficiently small, the distribution of $\rmX_{t_0}$ in the forward process is close enough to the real data distribution.
\end{remark}

\subsection{Proof of proposition \ref{prop:stat_error}}
\label{appendix: prop1 proof}
We first provide the following definition \ref{def:p_delta}, which is adopted from \citet{fu2024unveil}.
\begin{definition}
\label{def:p_delta}
    Denote the space of density functions 
\begin{equation*}
\mathcal{P}_0=\left\{p(\mX)=f(\mX)\exp(-C\Vert \mX\Vert_2^2): f\in\mathcal{L}(\R^{L\times H}, B), f(\mX)\geq \alpha>0\right\},
    \end{equation*}
where $C$ and $\alpha$ can be any given constants, and $\mathcal{L}(\R^{L\times H}, B)$ denotes the class of Lipschitz continuous functions on $\R^{L\times H}$ with Lipschitz constant bounded by $B$. 

Suppose that the density  function of $\rmX$ belongs to the following space 
\begin{equation}
    \mathcal{P}_\delta=\left\{ p(\mX)\in\mathcal{P}_0|P(\bar{R},O)=\delta\right\},
\end{equation}
where the distribution of $\rmM$ is defined from $\rmX$ by $$\ermM_{lh} = \vone\{\ermX_{lh}\geq 1/2\}.$$
\end{definition}
\begin{proposition}
\label{prop:stat_error_appendix}
    Consider generating piano roll $\rmM$ from a continuous random variable $\rmX$, i.e.,  given $n$ i.i.d. data $\{\rmX^{i}\}_{i=1}^n\sim p_\rmX$, let $\{\rmM^{i}\}_{i=1}^n$ be given by  $\ermM^i_{lh} = \vone\{\ermX^i_{lh}\geq 1/2\}$. Denote the model for estimating the distribution of $\mathbf{X}$ as $\widehat{p}_\rmX$. We have  $\exists~C>0$ such that $\forall n$,
\begin{equation}
    \label{eq:decay2}\inf_{\widehat{p}_\rmX}\sup_{p_\rmX\in\mathcal{P}_\delta}\E_{\{\rmM^i\}_{i=1}^n}\widehat{P}(\bar{R},O)\geq C\cdot
 n^{-\frac{1}{LH+2}}-P(\bar{R},O),
\end{equation}
where $\widehat{P}$ is the probability associated with the generated data $\estpx$.

\end{proposition}

\begin{proof}
We first restate a special case of proposition 4.3 of \citet{fu2024unveil} as the following lemma.
\begin{lemma}(\citet{fu2024unveil}, proposition 4.3)
    Fix a constant \( C_2 > 0 \). Consider estimating a distribution \( P(\mathbf{x}) \) with a density function belonging to the space
\[
\mathcal{P} = \left\{ p(\mathbf{x}) = f(\mathbf{x}) \exp(-C_2 \|\mathbf{x}\|_2^2) : f(\mathbf{x}) \in \mathcal{L}(\mathbb{R}^d, {B}), f(\mathbf{x}) \geq C > 0 \right\}.
\]
Given \( n \) i.i.d. data \( \{ x_i \}_{i=1}^n \), we have
\[
\inf_{\hat{\mu}} \sup_{p \in \mathcal{P}} \mathbb{E}_{\{x_i\}_{i=1}^n} \left[\text{TV}(\hat{\mu}, P)\right] \gtrsim n^{-\frac{1}{d + 2}},
\]
where the infimum is taken over all possible estimators \( \hat{\mu} \) based on the data.
\label{lemma:fu}
\end{lemma}

From lemma \ref{lemma:fu}, since the space $\mathcal{P}_0$ that we define satisfies all the same conditions as the space $\mathcal{P}$ in lemma \ref{lemma:fu}, we know from the conclusion of lemma \ref{lemma:fu}  that 
\begin{equation}
    \inf_{\estpx} \sup_{p_\rmX \in \mathcal{P}_0} \mathbb{E}_{\{x_i\}_{i=1}^n} \left[\text{TV}(\estpx, p_\rmX)\right] \gtrsim n^{-\frac{1}{LH + 2}},
    \label{eq:TV from lemma}
\end{equation}
where by definition of total variation,
\begin{equation}
    \text{TV}(\estpx, p_{\rmX})=\int_{\sR^{L\times H}}|\estpx(\mX)-p_\rmX(\mX)|d\mX.
    \label{eq:TV}
\end{equation}

For simplicity, suppose event $O$ denotes a note-out-of-key occurring at $(l,h)=(1,1)$.
We have 
\begin{equation}
    \label{eq:int1}
    \begin{aligned}\widehat{P}(O
)&=\int_{(\frac{1}{2},+\infty)} d\emX_{11}\int_{\sR^{L\times H-1}}d\mY~\estpx(\emX_{11},\mY)\\
    &\overset{\Delta}{=}\int_{\Omega_{O}}\estpx(\mX)d\mX,
    \end{aligned}
\end{equation}
where $\mY$ is a $(LH-1)$-dimensional variable denoting the elements in matrix $\mX$ excluding $\emX_{11}$. Let
$\sC(O)$ denote the set of all possible realizations of piano roll $\mM$ that contains (i) the note $O$ as an out-of-key note, and (ii) a ``resolution''\footnote{By definition, the resolution of an out-of-key note refers to how it is integrated into the surrounding harmonic and melodic structure to make it sound intentional rather than an error.} to accommodate it. For each $\mM\in\sC(O)$, let $$\delta(\mM)=\{(l,h)\in \llbracket 1,L\rrbracket\times \llbracket 1, H\rrbracket|\emM_{lh}=1\}.$$ 

Therefore, we have 
\begin{equation}
\begin{aligned}
\widehat{P}(R,O)&=\sum_{\mM\in\sC(O)}\int_{(\frac{1}{2},+\infty)^{|\delta(\mM)|}}d\emX_{\delta(\mM)}\int_{(-\infty,\frac{1}{2})^{L\times H-|\delta(\mM)|}}d\mY~\estpx(\emX_{\delta(\mM)},\emX_{L\times H\backslash\delta(\mM)})\\
&\overset{\Delta}{=}\int_{\Omega_{\sC(O)}}\estpx(\mX)d\mX,
\end{aligned}
\label{eq:int2}
\end{equation}
and note that $\Omega_{\sC(O)}\subset\Omega_{O}$, we have
\begin{equation}
    \widehat{P}(\bar{R},O)=\widehat{P}(O)-\widehat{P}(R,O) = \int_{\Omega_{O}\backslash\Omega_{\sC(O)}}\estpx(\mX)d\mX\label{eq:int3}
\end{equation}
To better explain and summarize \eqref{eq:int1}, \eqref{eq:int2} and \eqref{eq:int3}, the probabilities   $\widehat{P}(\cdot)$ (the estimated probabilities of $O,\{R,O\}$ or $\{\bar{R},O\}$) are always calculated from integrating $\estpx(\mX)$ on a corresponding domain, and the key of the 3 equations are all about finding the domain on which to integrate . Similarly, for the ground truth distributions and under definition \ref{def:p_delta} which provides $P_\mM(\bar{R},O)=\delta$, we have 
\begin{equation*}
    P(\bar{R},O)=\int_{\Omega_{O}\backslash\Omega_{\sC(O)}}p_\rmX(\mX)d\mX\leq \delta.
\end{equation*}
Therefore, 
\begin{equation}
   \begin{aligned} \widehat{P}(\bar{R},O)&=\int_{\Omega_{O}\backslash\Omega_{\sC(O)}}\estpx(\mX) d\mX\\
   &\geq \int_{\Omega_{O}\backslash\Omega_{\sC(O)}}\left|\estpx(\mX)-p_\rmX(\mX)\right|-p_\rmX(\mX)d\mX\\
   &\geq\int_{\Omega_{O}\backslash\Omega_{\sC(O)}}\left|\estpx(\mX)-p_\rmX(\mX)\right|d\mX-\delta
   \end{aligned}
\end{equation}
Therefore,
\begin{equation}
    \label{eq:P and TV}\widehat{P}(\bar{R},O)= \text{TV}|_{\Omega_{O}\backslash\Omega_{\sC(O)}}(\estpx,p_\rmX)-\delta,
\end{equation}
where $\text{TV}|_{\Omega_{O}\backslash\Omega_{\sC(O)}}$ is the total variation integral restricted on the domain $\Omega_{O}\backslash\Omega_{\sC(O)}$. 

By construction of packing numbers provided in the proof of proposition 4.3 of \citet{fu2024unveil}, we note that constraint $P_\mM(\bar{R},O)=\delta$ or restricting the integral of total variation on $\Omega_{O}\backslash\Omega_{\sC(O)}$ does not change the order of the packing numbers, i.e., $\mathcal{P}_0$ and $\mathcal{P}_\delta$ have the same packing numbers. Let 
$$\mathcal{P}_\delta^{\Omega_{O}\backslash\Omega_{\sC(O)}}=\left\{ C(\Omega_{O}\backslash\Omega_{\sC(O)})\cdot p(\mathbf{X})\mathbf{1}_{\mathbf{X}\in \Omega_{O}\backslash\Omega_{\sC(O)}}~|~p(\mathbf{X})\in\mathcal{P}_\delta \right\},$$ 
where the constant $C(\Omega_{O}\backslash\Omega_{\sC(O)})$ is a scale factor to ensure that $C(\Omega_{O}\backslash\Omega_{\sC(O)})\cdot p(\mathbf{X})\mathbf{1}_{\mathbf{X}\in \Omega_{O}\backslash\Omega_{\sC(O)}}$ is a probability density function. For simplicity we use $\mathcal{P}(\delta, O)$ for short of $\mathcal{P}_\delta^{\Omega_{O}\backslash\Omega_{\sC(O)}}$. 
Therefore, from the original lemma 1 of \citet{fu2024unveil} we have \eqref{eq:TV from lemma}. Only changing the $\mathcal{P}_0$ into $\mathcal{P}(\delta,O)$ (all the arguments above are to justify why this change can be made), we have
\begin{equation}
\label{eq:p_delta_O}
    \inf_{\estpx} \sup_{p \in \mathcal{P}(\delta,O)} \mathbb{E}_{\{\rmX_i\}_{i=1}^n}\text{TV}(\estpx,p_\rmX)\gtrsim n^{-\frac{1}{LH + 2}}.
\end{equation}
Combining \eqref{eq:p_delta_O} with \eqref{eq:P and TV}, and starting from our target $\mathbb{E}\widehat{P}(\bar{R},O)$,  we have 
\begin{equation*}
\begin{aligned}
    &\inf_{\estpx} \sup_{p \in \mathcal{P}_\delta} \mathbb{E}_{\{\rmX_i\}_{i=1}^n}\widehat{P}(\bar{R},O)+\delta=\inf_{\estpx} \sup_{p \in \mathcal{P}_\delta}\text{TV}|_{\Omega_{O}\backslash\Omega_{\sC(O)}}(\estpx,p_\rmX)+\delta\\
    =&\inf_{\estpx} \sup_{p \in \mathcal{P}(\delta,O)}\text{TV}(\estpx,p_\rmX) +\delta\gtrsim n^{-\frac{1}{LH + 2}}.
\end{aligned}
\end{equation*}
Therefore, $\exists C>0$, $\forall n$,
\begin{equation*}
    \inf_{\estpx} \sup_{p \in \mathcal{P}_\delta} \mathbb{E}_{\{\rmX_i\}_{i=1}^n}\widehat{P}(\bar{R},O)\geq C\cdot n^{-\frac{1}{LH + 2}}-P(\bar{R},O).
\end{equation*}
which finishes the proof.
\end{proof}

\section{Details of Conditioning and Algorithms}
\label{appendix:additional}

\subsection{Mathematical formulation of textural conditions in section \ref{subsec:fine-grained condition}}

Denote a chord progression by $\mathcal{C}$, where $\mathcal{C}(l)$ denotes the chord at time $l\in\llbracket 1, L\rrbracket$. Let $\gamma_{\mathcal{C}}(l)\subset \llbracket 1, H\rrbracket$ denote the set of pitch index $h$ that belongs to the pitch classes included in chord $\mathcal{C}(l).$\footnote{For example, when $\mathcal{C}(l)=$ C major (consisting of pitch classes C, E and G), $\gamma_\mathcal{C}$ includes all pitch values corresponding to the three pitch classes across all octaves. }, and let $\gamma_\mathcal{R}\subset \llbracket 1,L\rrbracket$ denote the set of onset time indexes corresponding to rhythmic pattern $\mathcal{R}$. We define the following versions of representations for the condition: 
\begin{itemize}
    \item When harmonic ($\mathcal{C}$) and rhythmic ($\mathcal{R}$) conditions are both provided, the corresponding conditional piano roll $\mMc(\idC,\idR)$ is given element-wise by $\emMc_{lh}(\idC,\idR)=\vone\{l\in\gamma_\idR\}\vone\{h\in\gamma_\idC(l)\}$, meaning that the $(l,h)$-element is $1$ if pitch index $h$ belongs to chord $\mathcal{C}(l)$ and there is onset notes at time $l$, and $0$ otherwise.
    \item When only harmonic ($\mathcal{C}$) condition is provided, the corresponding piano roll $\mMc(\idC)$ is given element-wise by $\emMc_{lh}(\idC)=-1-\vone\{h\in\gamma_\idC(l)\}$, meaning that the $(l,h)$-element is $-2$ if pitch index $h$ belongs to chord $\mathcal{C}(l)$, and $-1$ otherwise.
\end{itemize}
Figure~\ref{fig:McondCR} and Figure~\ref{fig:McondC} provides illustrative examples of $\mMc(\idC,\idR)$ and $\mMc(\idC)$. The use of $-2$ and $-1$ (rather than $1$ and $0$) in the latter case ensures that the model can fully capture the distinctions between the two scenarios, as a unified model will be trained on both types of conditions.

\subsection{Classifier Free Guidance}

To enable the model to generate under varying levels of conditioning, including unconditional generation, we implement the idea of classifier-free guidance, and randomly apply conditions with or without rhythmic pattern in the process of training. Namely, the training loss is modified from \eqref{eq:loss function} and given as
\begin{equation}
\label{eq:loss function_cond}  
\begin{aligned}
    \mathbb{E}_{t,{\vep}, \rmX_0}\left[\lambda_1(t)\Vert {\vep}-{\vep}_\theta(\rmX_t, \rmMc(\mathcal{C}), t)\Vert^2\right.\\
    \left.+\lambda_2(t)\Vert {\vep}-{\vep}_\theta(\rmX_t, \rmMc(\mathcal{C},\mathcal{R}), t)\Vert^2\right],
\end{aligned}
\end{equation}
where $\lambda_1(t)$ and $\lambda_2(t)$ are hyper-parameters. Note that both $\rmMc(\idC)$ and $\rmMc(\idC,\idR)$ are derived from $\rmX_0$ via pre-designed chord recognition and rhythmic identification algorithms.

The guided noise prediction  at timestep 
$t$ is then computed as
\begin{equation}
    \begin{aligned}{\vep}_\theta(\rmX_t,t|\idC,\idR)= &{\vep}_\theta(\rmX_t, \mMc(\mathcal{C}), t)\\
    &+w\cdot\left[{\vep}_\theta(\rmX_t, \mMc(\mathcal{C},\idR),t)\right.\\
    &\left. -{\vep}_\theta(\rmX_t, \mMc(\mathcal{C}),t)\right],
    \end{aligned}
    \label{eq:guided noise}
\end{equation}
where $w$ is the weight parameter. Note that the general formulation ${\vep}_\theta(\rmX_t,t|\idC,\idR)$ includes the case where rhythmic guidance is not provided ($\idR=\emptyset$), and $w$ in \eqref{eq:guided noise} is set as $0$.

\subsection{Additional algorithms in section \ref{subsec:fine-grained sampling}}
In this section, we provide the following algorithm: fine-grained sampling guidance additionally with rhythmic regularization, fine-grained sampling guidance combined with DDIM sampling.

Let $\mathcal{B}$ denote the rhythmic regularization. Specifically, we have the following types of regularization:
\begin{itemize}
    \item $\gB_1$: Requiring exactly $N$ onset of a note at time position $l$, i.e., $\sum_{h\in\llbracket 1,H\rrbracket}\emM_{lh}=N$
    \item $\gB_2$: Requiring at least $N$ onsets at time position $l$, i.e., $$\exists \vh\subset\llbracket 1,H\rrbracket,\text{ or }\exists \vh\subset \llbracket 1,H\rrbracket\backslash\omega_\idK(l) \text{ if harmonic regularization is jointly included}$$ such that $\emM_{l\vh}=1$, and $|\vh|\geq N$
    \item $\gB_3$: Requiring no onset of notes at time position $l$, i.e., $\forall h\in\llbracket 1,H\rrbracket$, $\emM_{lh}=0$
\end{itemize}

Let the set of $\mM$ satisfying a specific regularization $\mathcal{B}$ be denoted as $\sM_\gB$, and the corresponding set of $\mX$ be denoted as $\tilde{\sM}_\gB$, note that this includes the case where multiple requirements are satisfied, resulting in $$\tilde{\sM}_\gB=\tilde{\sM}_{\gB_1,\gB_2,\ldots}=\tilde{\sM}_{\gB_1}\cap\tilde{\sM}_{\gB_2}\cap\ldots.$$
The correction of predicted noise score is then formulated as 
\begin{equation}
\begin{aligned}
    \tilde{\vep}_\theta(\mX_t,t|\idC,\idR) = \argmin_\vep \quad & \Vert \vep- \widehat{\vep}_\theta(\mX_t,t|\idC,\idR)\Vert\\
    \text{ s.t. } \quad & \left(\frac{\mX_t-\sqrt{1-\bar{\alpha}_t}\vep}{\sqrt{\bar{\alpha}_t}}\right)\in \tilde{\sM}_\gB.
\end{aligned}
\label{eq:optimization problem_rhy}
\end{equation}
Further, we can perform predicted noise score correction with joint regularization on rhythm and harmony, resulting in the corrected noise score 
\begin{equation}
\begin{aligned}
    \tilde{\vep}_\theta(\mX_t,t|\idC,\idR) = \argmin_\vep \quad & \Vert \vep- \widehat{\vep}_\theta(\mX_t,t|\idC,\idR)\Vert\\
    \text{ s.t. } \quad & \left(\frac{\mX_t-\sqrt{1-\bar{\alpha}_t}\vep}{\sqrt{\bar{\alpha}_t}}\right)\in(\sR^{L\times H}\backslash \sW^\prime_\idK)\cap \tilde{\sM}_\gB.
\end{aligned}
\label{eq:optimization problem_joint}
\end{equation}

We for example provide a element-wise solution 
of $\tilde{\vep}_\theta(\mX_t,t|\idC,\idR)$ defined by problem (\ref{eq:optimization problem_rhy}). For given $l$, suppose $\mathcal{B}(l)$ takes the form of $\mathcal{B}_2$, for simplicity take $N=1$. This gives $\tilde{\vep}_{\theta,lh}=\widehat{\vep}_{\theta,lh}$ if
$\max_{h}\E[\rmX_0|\mX_t]_{hl}\geq \frac{1}{2}$
and 
$\E[\rmX_0|\mX_t]_{hl}=\frac{1}{2}$, $h=\arg\max_{h}\E[\rmX_0|\mX_t]_{hl},$
i.e.,
$$\tilde{\vep}_{\theta,lh}=\frac{1}{\sqrt{1-\bar{\alpha}_t}}\left(\emX_{t,lh}-\frac{\sqrt{\bar{\alpha}_t}}{2}\right),$$
if $\max_{h}\E[\rmX_0|\mX_t]_{hl}< \frac{1}{2}$. The correction applied to predicted $\rmX_0$ ($\E[\rmX_0|\mX_t]$) is illustrated in the following figure \ref{fig:illustrative example of control X0}.

\begin{figure}[htbp]
    \centering
    
    \begin{subfigure}[b]{0.6\linewidth}
        \centering
        \includegraphics[width=\textwidth]{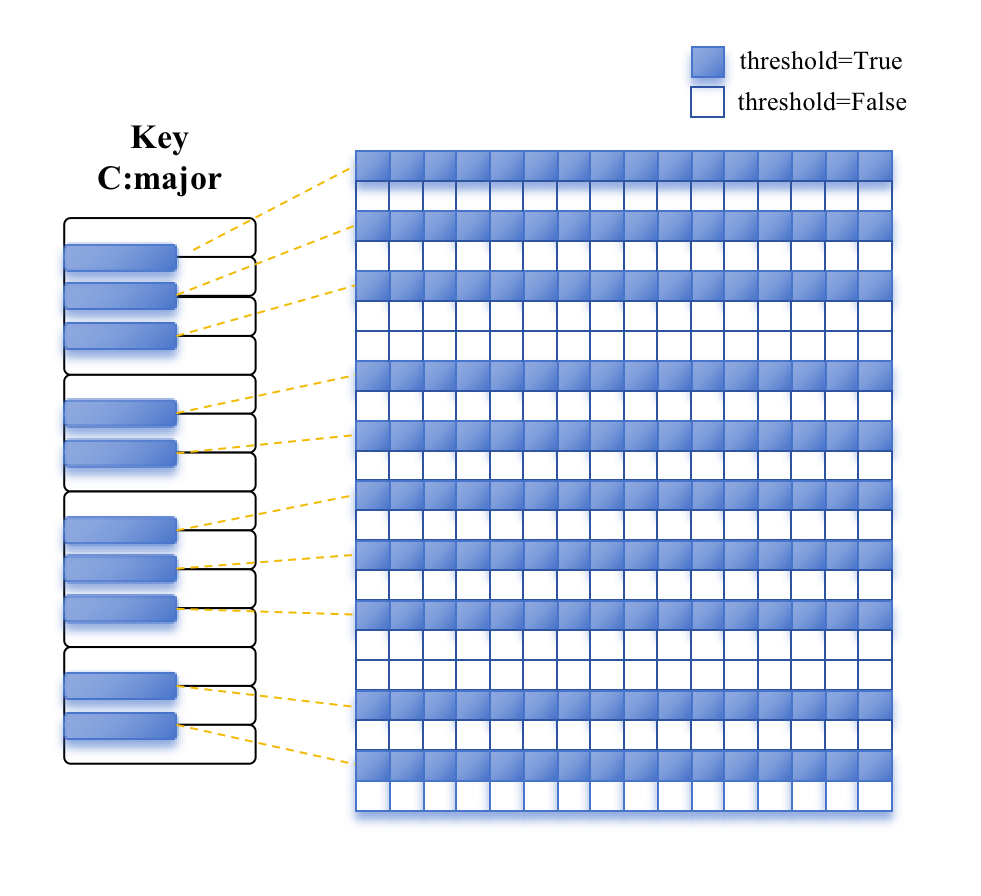}  
        \caption{Fine-grained control for $\E[\rmX_0|\mX_t]\in \sR^{L\times H}\backslash\sW^\prime_\idK$. The colored spots denote places that we require $\E[\rmX_0|\mX_t]_{lh}\leq \frac{1}{2}$.}
        \label{fig:harmonic control}
    \end{subfigure}
    \begin{subfigure}[b]{0.6\linewidth}
        \centering
        \includegraphics[width=\textwidth]{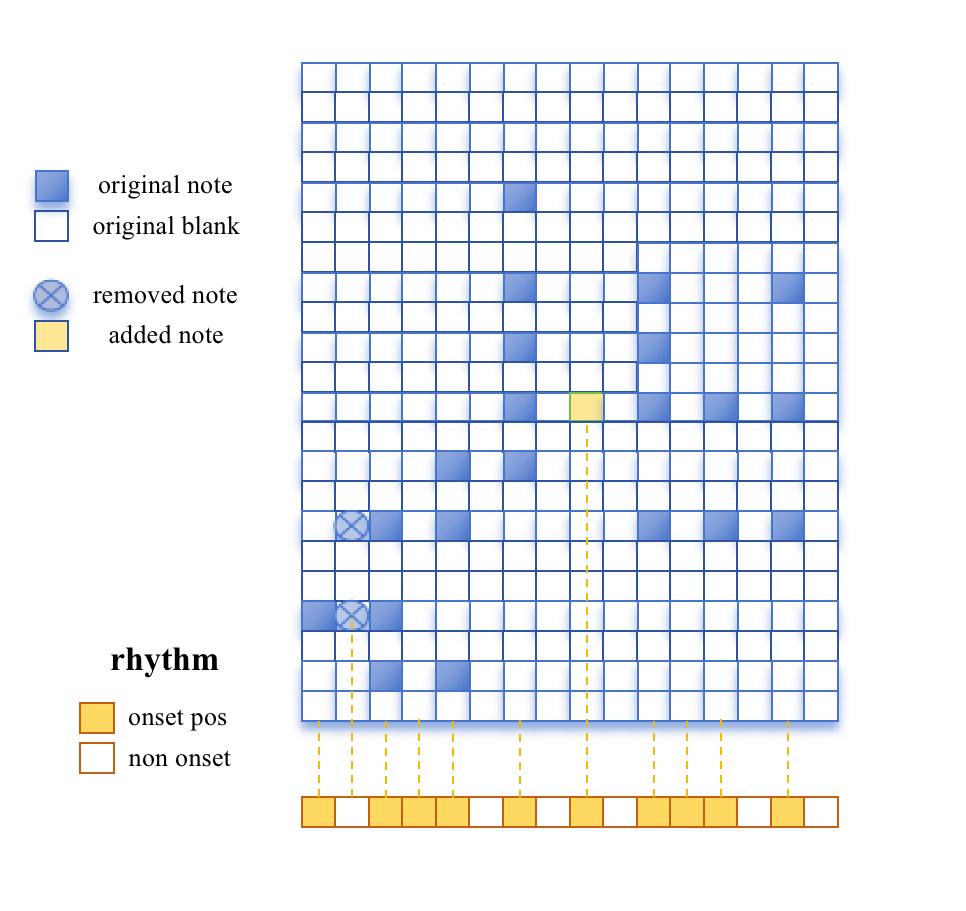}  
        \caption{Fine-grained control for $\E[\rmX_0|\mX_t]\in \sW^\prime_\gB$. Original notes are removed at $l$ if $\gB_3$ is applied. Otherwise if $\gB_1$ is applied and currently no note exists, the ``most likely notes'' (i.e., at $h=\arg\max\E[\rmX_0|\mX_t]_{lh}$) are added.}
        \label{fig:rhythm control}
    \end{subfigure}
    
    \caption{Illustration of fine-grained control on predicted $\rmX_0$.}
    \label{fig:illustrative example of control X0}
\end{figure}

\begin{algorithm}[H]
\label{alg:sample_rhy}
\SetAlgoLined
\KwIn{Input parameters: forward process variances $\beta_t$, $\bar{\alpha}_t=\prod_{s=1}^t\beta_t$, backward noise scale $\sigma_t$, chord condition $\idC$, key signature $\idK$, rhythmic condition $\idR$, rhythmic guidance $\gB$ }
\KwOut{generated piano roll $\tilde{\rmM}\in\{0,1\}^{L\times H}$}
$\rmX_T\sim\mathcal{N}(0,\mI)$;

\For{$t = T,T-1,\ldots,1$}{
    Compute guided noise prediction $\widehat{\vep}_\theta(\mX_t,t|\idC,\idR)$\;
    
    Perform noise correction:   derive $\tilde{\vep}_\theta(\mX_t,t|\idC,\idR)$ optimization \eqref{eq:optimization problem_joint}\;

    Compute $\tilde{\rmX}_{t-1}$ by plugging the corrected noise $\tilde{\vep}_\theta(\mX_t,t|\idC,\idR)$ into \eqref{eq: DDIM}
}
Convert $\tilde{\rmX}_0$ into piano roll $\tilde{\rmM}$ 

\Return{$output$}\;
\caption{DDPM sampling with fine-grained textural guidance}
\end{algorithm}

We additionally remark that the fine-grained sampling guidance is empirically effective with the DDIM sampling scheme, which drastically improves the generation speed. Specifically, select subset $\{\tau_i\}_{i=1}^m\subset\llbracket 1,T\rrbracket$, and denote
\begin{equation*}
   \rmX_{\tau_{i-1}}=\sqrt{\bar{\alpha}_{\tau_{i-1}}}\left(\frac{\rmX_t-\sqrt{1-\bar{\alpha}_{\tau_i}}\widehat{\vep}_\theta(\rmX_{\tau_i},\tau_i)}{\sqrt{\bar{\alpha}_{\tau_i}}}\right) +\sqrt{1-\bar{\alpha}_{\tau_{i-1}}-\sigma_{\tau_i}^2}\widehat{\vep}_\theta(\rmX_{\tau_i},\tau_i)+\sigma_{\tau_i}\vep_{\tau_i},
\end{equation*}
we similarly perform the DDIM noise correction
\begin{equation*}
\begin{aligned}
    \tilde{\vep}_\theta(\mX_{\tau_i},\tau_i|\idC,\idR) = \argmin_\vep \quad & \Vert \vep- \widehat{\vep}_\theta(\mX_{\tau_i},\tau_i|\idC,\idR)\Vert\\
    \text{ s.t. } \quad & \left(\frac{\mX_t-\sqrt{1-\bar{\alpha}_{\tau_i}}\vep}{\sqrt{\bar{\alpha}_{\tau_i}}}\right)\in(\sR^{L\times H}\backslash \sW^\prime_\idK)\cap \tilde{\sM}_\gB.
\end{aligned}
\end{equation*}
on each step $i$.

\section{Comparison with Related Works}
\label{appendix: comparison}
We provide a detailed comparison between our method and two related works in controlled diffusion models with constrained or guided intermediate sampling steps:

\textbf{Comparison with reflected diffusion models}  In \citet{lou2023reflected}, a bounded setting is used for both the forward and backward processes, ensuring that the bound applies to the training objective as well as the entire sampling process. In contrast, we do not adopt the framework of bounded Brownian motion, because we do not require the entire sampling process to be bounded within a given domain; instead, we only enforce that the final sample outcome aligns with the constraint. While \citet{lou2023reflected} enforces thresholding on $\rmX_t$ in both forward and backward processes, our approach is to perform a thresholding-like projection method on the predicted noise $\vep_\theta(\rmX_t,t)$, interpreted as noise correction.

\textbf{Comparison with non-differentiable rule guided diffusion} \citet{huang2024symbolic} guides the output with musical rules by sampling multiple times at intermediate steps, and continuing with the sample that best fits the musical rule, producing high-quality, rule-guided music. Our work centers on a different aspect,  prioritizing precise control to tackle the challenges of accuracy and regularization in symbolic music generation. Also, we place additional emphasis on sampling speed, ensuring stable generation of samples within seconds to facilitate interactive music creation and improvisation.

\section{Numerical Experiment Details}
\subsection{Detailed Data Representation}
\label{appendix:data representation details}

The two-channel version of piano roll with with both harmonic and rhythm conditions ($\rmMc(\mathcal{C},\mathcal{R})$) and with harmonic condition ($\rmMc(\mathcal{C})$) with onset and sustain are represented as:
\begin{itemize}
    \item $\rmMc(\mathcal{C},\mathcal{R})$: In the first channel, the $(l,h)$-element is $1$ if there are onset notes at time $l$ and pitch index $h$ belongs to the chord $\mathcal{C}(l)$, and $0$ otherwise. In the second channel, the $(l,h)$-element is $1$ if pitch index $h$ belongs to the chord $\mathcal{C}(l)$ and there is no onset note at time $l$.
    \item $\rmMc(\mathcal{C})$: In both channels, the $(l,h)$-element is $1$ if pitch index $h$ belongs to the chord $\mathcal{C}(l)$, and $0$ otherwise.
\end{itemize}

In each diffusion step $t$, the model input is a concatenated 4-channel piano roll with shape $4\times L\times 128$, where the first two channels correspond to the noisy target $\mX_t$ and the last two channels correspond to the condition $\mMc$ (either $\rmMc(\mathcal{C},\mathcal{R})$ or $\rmMc(\mathcal{C})$). The output is the noise prediction $\hat{\varepsilon_\theta}$, which is a 2-channel piano roll with the same shape as $\mX_t$. 
For the accompaniment generation experiments, we provide melody as an additional condition, which is also represented by a 2-channel piano roll with shape $2\times L\times 128$, with the same resolution and length as $\mX$. The melody condition is also concatenated with $\mX_t$ and $\mMc$ as model input, which results in a full 6-channel matrix with shape $6\times L\times 128$.

\subsection{Training and Sampling Details}
\label{appendix:training and sampling details}
We set diffusion timesteps $T=1000$ with $\beta_0=8.5e{-4}$ and $\beta_T=1.2e{-2}$. We use AdamW optimizer with a learning rate of $5e{-5}$, $\beta_1=0.9$, and $\beta_2=0.999$. We applied data augmentation by transposing each 4-measure piece into all 12 keys. This involves uniformly shifting the pitch of all notes and adjusting the corresponding chords accordingly. This augmentation expands the dataset to 189,132 samples. Training is conducted with a batch size of 16, utilizing random sampling without replacement. Specifically, in each iteration, 16 samples are randomly selected without replacement until all samples are utilized, constituting one epoch. This procedure is repeated to ensure each sample was processed twice during training, resulting in a total of 23,642 iterations.

To speed up the sampling process, we select a sub-sequence of length 10 from $\{1,\cdots,T\}$ and apply the accelerated sampling process in \citet{song2020denoising}. It takes 0.4 seconds to generate the 4-measure accompaniment on a NVIDIA RTX 6000 Ada Generation GPU.

\subsection{Experiments on Symbolic Music Generation Given only Chord Conditions}
\label{appendix:music generation task}
As mentioned in Section~\ref{subsec:experiment}, we also run numerical experiments on symbolic music generation tasks given only chord condition. However, compared with the accompaniment generation task, we remark that this experiment does not have enough effective basis for comparison.

For the accompaniment generation task, we evaluate the cosine similarity of chord progression between the generated samples and the ground truth, as well as the IoU of chord and piano roll. The comparison with ground truth on those features make sense in the accompaniment generation task, because the leading melody inherently contains many constraints on the rhythm and pitch range of the accompaniment, ensuring coherence with the melody. Thus, similarity with ground truth on those metrics serves as an indicator of how well the generated samples adhere to the melody.

However, in symbolic music generation conditioned only on a chord sequence, while chord progression similarity remains comparable (as the chord sequence is provided), evaluating IoU of piano roll against ground truth is less informative. This is because multiple different pitch range and rhythm could appropriately align with a given chord progression, making deviations from the ground truth in these features less indicative of sample quality. Therefore, chord similarity emerges as the sole applicable metric in this context.

Additionally, WholeSongGen's architecture does not support music generation conditioned solely on chord progressions, as it utilizes a shared piano-roll for both chord and melody, rendering it unsuitable for comparison. Conversely, GETMusic facilitates the generation of both melody and piano accompaniment based on chord conditions, allowing for a viable comparison.

Consequently, we present results focusing on chord similarity between our model and GETMusic. For our model, we evaluate performance under two conditions: with both conditioning and control during training and sampling, and with conditioning during training but without control during sampling. The outcomes, summarized in Table~\ref{tab:evaluation-table-no-mel}, indicate that our fully controlled FGG method surpasses both the one without sampling control and GETMusic.

\begin{table}[ht]
\begin{center}
\renewcommand{\arraystretch}{1.2} 
\setlength{\tabcolsep}{10pt} 
\small 
\begin{tabular}{l|llll}
\hline
\rowcolor[HTML]{D9EAD3}  
\multicolumn{1}{c|}{\bf Methods}  &\multicolumn{1}{c}{\bf FGG (Ours)} & \multicolumn{1}{c}{FGG, only Training control} & \multicolumn{1}{c}{GETMusic} \\
\hline
\rowcolor[HTML]{F4F4F4}  
\textbf{Chord Similarity} & $\bm{0.676\pm 0.007}$ & ${0.645\pm 0.008}$ & $0.499\pm 0.013$\\
\hline
\end{tabular}
\end{center}
\caption{Evaluation of the similarity with ground truth, chord-conditioned music generation.}
\label{tab:evaluation-table-no-mel}
\end{table}

\section{Demo Page Details}
In this section, we briefly introduce how the Dorian mode and Chinese style clips are generated. We note that both styles are shaped not only by key-constraint $\mathcal{K}$, but also with designed chord progressions $\mathcal{C}$. 

The key constraint for Dorian mode, example 1, is $\mathcal{K}_1=\{A,B,C,D,E,F\#,G\}$ throughout the 4 bars, which means all generated notes have to be in the pitch classes in $\mathcal{K}_1$. The the chord progression for Dorian mode, example 1 is $$\mathcal{C}_1=\text{Am} (4)- \text{Em} (2)-\text{Am}(2) - \text{C}(2)-\text{D}(2)-\text{Am}(2)-\text{D}(2).$$
For example 2, $\mathcal{K}_2=\{D, E, F, G, A, B, C\}$, and
$$\mathcal{C}_2=\text{Dm} (4)- \text{G} (4)- \text{C}(4)-\text{F}(4).$$

The number in parentheses corresponds to the number of beats the chord lasts. For example, at the beginning of $\mathcal{C}_1$, the chord Am lasts 4 beats. Therefore, for the condition matrix under the 16th resolution, the positions corresponding to pitch classes A, C and E have value 1, where the rest have value 0, for $t=0,1,2,\ldots, 15$. The condition is passed to the diffusion model as generation condition. Then $\mathcal{K}_1$ is applied as sampling control to shape and refine the tonal quality.

Similarly, for Chinese mode, we have $\mathcal{K}_1=\{C,D,E,G,A\}$ and 
$$\mathcal{C}_1=\text{G}(2)-\text{Am}(2)-\text{C}(2)-\text{G}(2)-\text{Em}(2)-\text{G}(2)-\text{D}(4).$$
For the second example, $\mathcal{K}_2=\{ D,E,\# F, A,  B\}$, and 
$$\mathcal{C}_2 = \text{A}(4)-\text{Bm}(2)\text{Fm}(2)-\text{Bm}(2)-\text{A}(2)-\text{Fm}(2)-\text{A}(2).$$

\section{Subjective Evaluation}
\label{appendix:subjective}
To compare performance of our FGG method against the baselines (WholeSongGen and GETMusic), we prepared 6 sets of generated samples, with each set containing the melody paired with accompaniments generated by FGG, WholeSongGen, and GETMusic, along with the ground truth accompaniment. This yields a total of $6\times 4 = 24$ samples. The samples are presented in a randomized order, and their sources are not disclosed to participants. Experienced listeners assess the quality of samples
in 5 dimensions: creativity, harmony (whether the accompaniment is in harmony with the melody),
melodiousness, naturalness and richness, together with an overall assessment.

\subsection{Background of Participants}
To evaluate the musical background of the participants, we first present the following questions:
\begin{itemize}
\item How many instruments (including vocal) are you playing or have you played?

\item Please list all instruments (including vocal) that you are playing or have played.

\item What is the instrument (including vocal) you have played the longest, and how many years have you been playing it? (e.g., piano, 3 years)
\end{itemize}

We recruited 31 participants with substantial musical experience for our survey. The number of instruments these participants play range from $0$ to $5$, with an average value of $2.03$, and a standard deviation of $1.31$. Examples of instrument played include piano, violin, vocal, guitar, saxphone, Dizi, Yangqin and Guzheng. The average years of playing has an average of $8.61$ and standard deviation of $8.08$. Specifically, the percentage of participants with $\geq 3$ years of playing music is $67.74\%$, and the percentage of participants with $\geq 10$ years of playing music is $45.16\%$. The distributions are given in the following figure \ref{fig:participants}.
\begin{figure}[ht]
    \centering
    
    \begin{subfigure}[b]{0.55\linewidth}
        \centering
        \includegraphics[width=\textwidth]{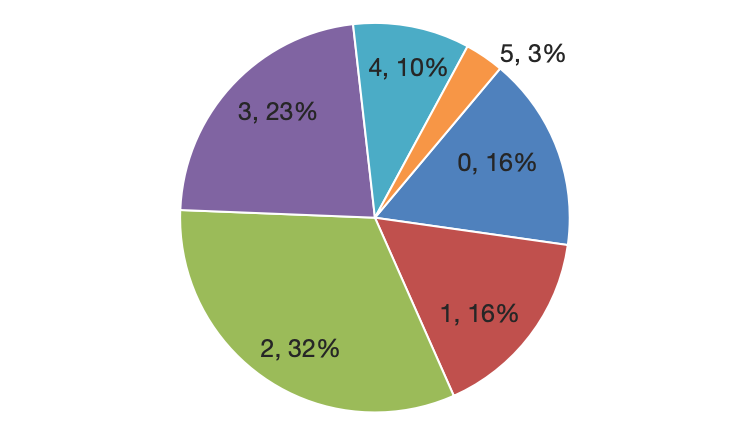}  
        \caption{Number of instruments played by the participants.}
        \label{fig:number}
    \end{subfigure}
    \begin{subfigure}[b]{0.65\linewidth}
        \centering
        \includegraphics[width=\textwidth]{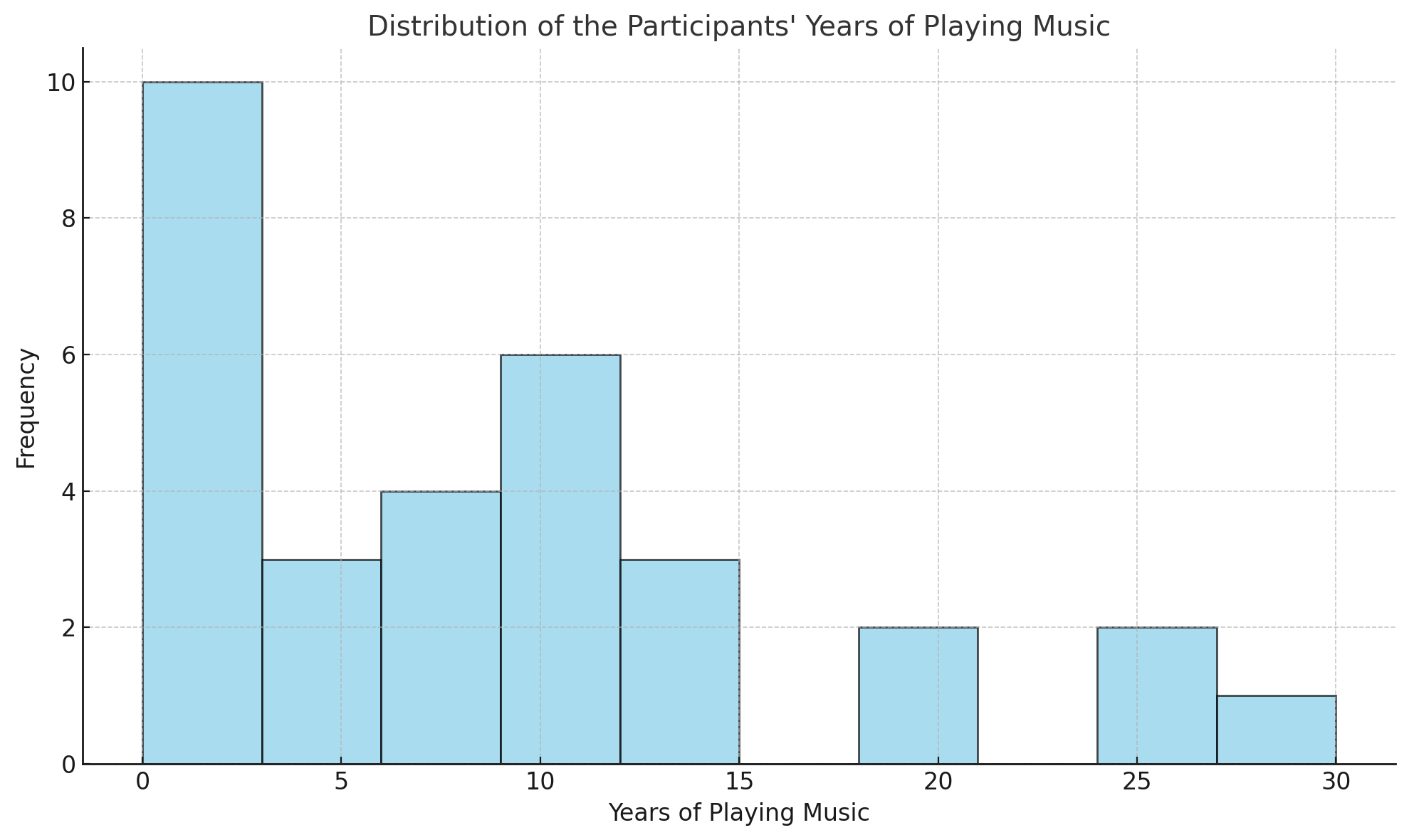}  
        \caption{Distribution of the participants' years of playing instruments.}
        \label{fig:years}
    \end{subfigure}
    
    \caption{Information of the musical background of the participants in the subjective evaluation.}
    \label{fig:participants}
\end{figure}

\subsection{Evaluation Questions}
Thank you for taking the time to participate in this experiment. You will be presented with 6 sets of clips, each containing 4 clips. The first clip in each set features the melody alone, while the remaining three include the melody accompanied by different accompaniments. After listening to each clip, please evaluate the accompaniments in the following dimensions based on your own experience.
\begin{itemize}
\item Does the accompaniment sound pleasant to you?
\item How would you rate the richness (i.e., the complexity, fullness, and expressive depth) of the accompaniment?
\item Does the accompaniment sound natural to you?
\item Do you think the accompaniment aligns well with the melody?
\item Does the accompaniment sound creative to you?
\item Please give an overall score for the clip.
\end{itemize}
For each question, participants are provided with a Likert scale ranging from 
1 to 5, where 
1 represents ``very poor'' and 5 represents ``very good.''

\section{Representative Examples of Sampling Control}
\label{appendix:empirical}

\begin{figure}[ht]
    \centering
\begin{subfigure}[b]{0.45\textwidth}
        \centering
\includegraphics[width=\linewidth]{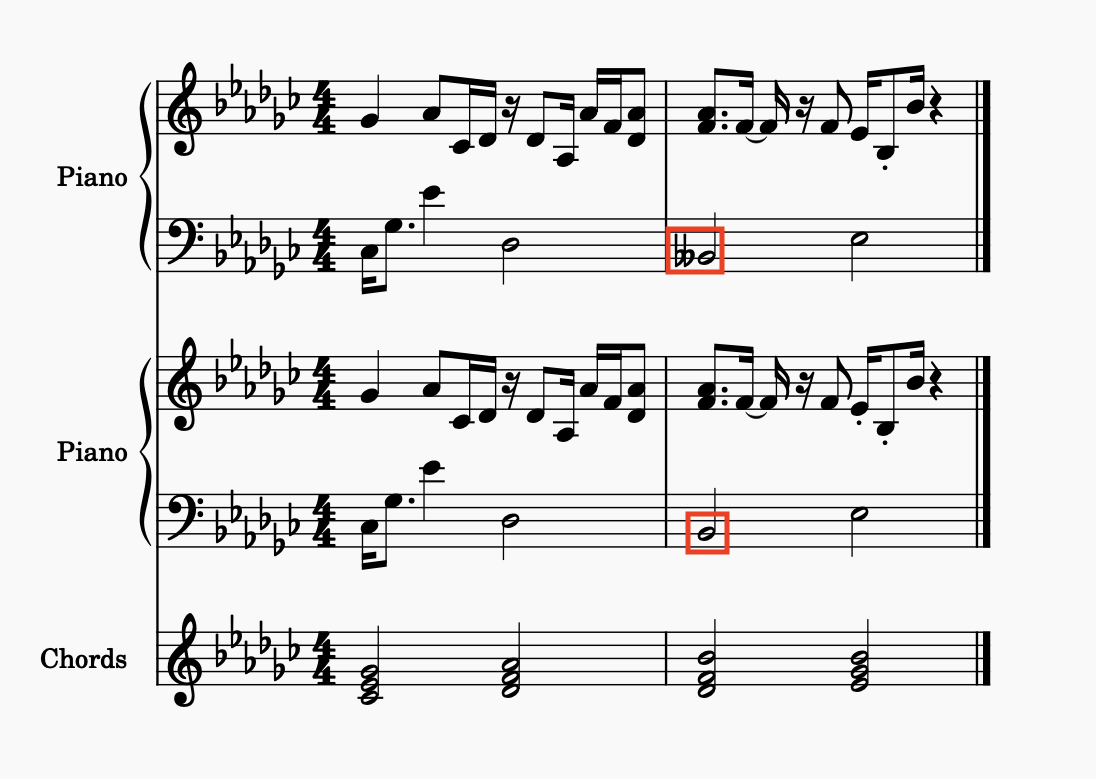}

\caption{An example of replacing an out-of-key note B$\flat\flat$ with the in-key note B$\flat$.}
        \label{fig:subfig1}
    \end{subfigure}
    \hspace{0.05\textwidth}
\begin{subfigure}[b]{0.45\textwidth}
        \centering
        \includegraphics[width=0.95\linewidth]{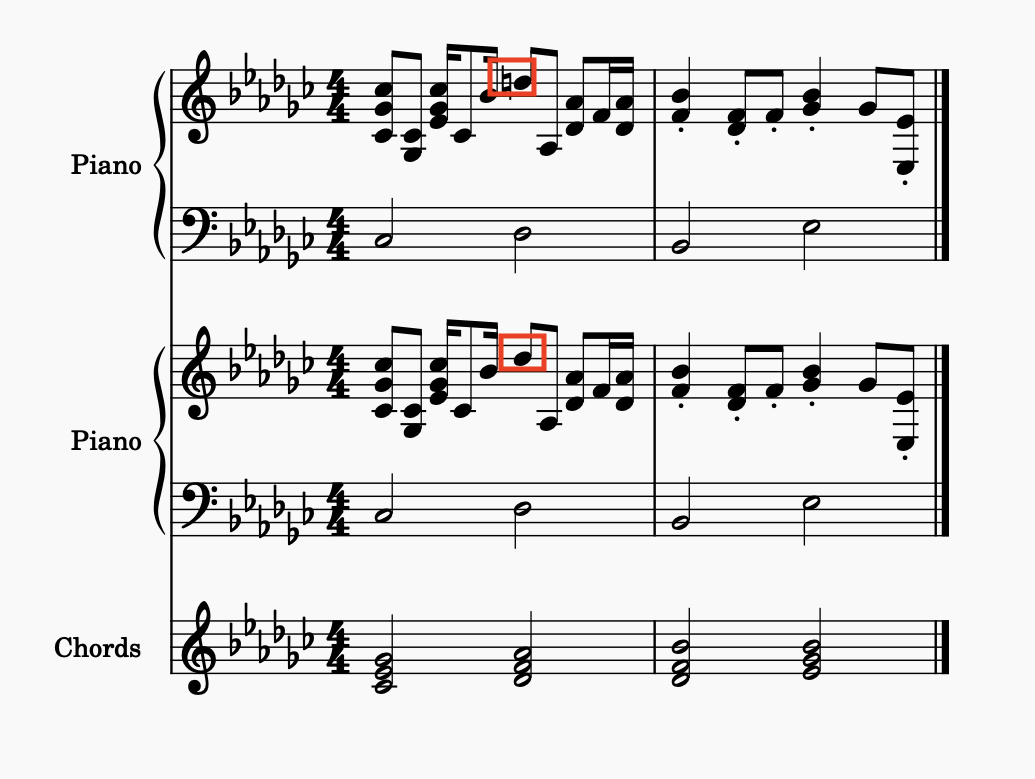}  
\caption{An example of replacing an out-of-key note D$\natural$ with the in-key note D$\flat$.}
        \label{fig:subfig2}
\end{subfigure}

\caption{Examples resulting from symbolic music generation with FGG. The first track is generated without key-signature control in sampling, the second track is generated with key-signature sampling control. The third track presents the chord condition. In each subfigure, the tracks are generated with the same conditions and the same set of noise. }
\label{fig:replace}
\end{figure}

In this section, we provide empirical examples of how model output is reshaped by fine-grained correction in Figure \ref{fig:replace}.  Notably, harmonic control not only helps the model eliminate incorrect notes, but also guides it to replace them with correct ones.

\section{The Effect of Guidance Weight for Classifier-free Guidance}
In Section~\ref{subsec:fine-grained condition}, we discussed the implementation of classifier-free guidance for rhythmic patterns, designed to enable the model to generate outputs under varying levels of conditioning. Specifically, we randomly apply conditions with or without rhythmic patter in the process of training. This approach ensures that the model can function effectively with both chord and rhythmic conditions or with chord conditions alone. Following \citet{ho2021classifier}, when generating with both chord and rhythmic conditions, the guided noise prediction at timestep 
$t$ is computed as:
\begin{equation*}
    \begin{aligned}{\vep}_\theta(\rmX_t,t|\idC,\idR)=& {\vep}_\theta(\rmX_t, \mMc(\mathcal{C}), t)\\
    &+w\cdot\left[{\vep}_\theta(\rmX_t, \mMc(\mathcal{C},\idR),t)-{\vep}_\theta(\rmX_t, \mMc(\mathcal{C}),t)\right],
    \end{aligned}
\end{equation*}
where ${\vep}_\theta(\rmX_t, \mMc(\mathcal{C},\idR),t)$ is the model's predicted noise without rhythmic condition, and ${\vep}_\theta(\rmX_t, \mMc(\mathcal{C},\idR),t)$ is the model's predicted noise with rhythmic condition, and $w$ is the guidance weight.

The literature has consistently demonstrated that the guidance weight $w$ plays a pivotal role in balancing diversity and stability in generation tasks \citep{ho2021classifier, chang2023muse, gao2023masked, lin2024common}. In general, a lower weight $w$ enhances sample diversity and quality, but this may come at the cost of deviation from the provided conditions. Conversely, higher values of $w$ promote closer adherence to the conditioning input, but excessively high $w$ can degrade output quality by over-constraining the model, resulting in less natural or lower-quality samples. 

In this section, we hope to investigate the effect of the guidance weight $w$ on our music generation task. We focus on the same accompaniment generation task as mentioned in Section~\ref{sec:experiments}. To measure the samples' adherence to rhythmic controls, we use the rhythm of the ground truth as the rhythmic condition and assess the overlapping area (OA) of note duration and note density between the generated and ground-truth samples. Specifically, we split both the generated accompaniments and the ground truth into non-overlapping 2-measure segments. Following \cite{von2023figaro}, for each feature $f$ ($f\in\{\text{note duration}, \text{note density}\}$), we calculate the macro overlapping area (MOA) in segment-level feature distributions so that the metric also considers the temporal order of the features. MOA is defined as
\begin{equation*}
    MOA(f) = \frac{1}{N}\sum_{i=1}^N \text{overlap}(\pi_i^{\text{gen}}(f), \pi_i^{\text{gt}}(f)),
\end{equation*}
where $\pi_i^{\text{gen}}(f)$ is the distribution of feature $f$ in the $i$-th generated segment, and $\pi_i^{\text{gt}}(f))$ is the distribution of feature $f$ in the $i$-th ground truth segment. Additionally, we measured the percentage of out-of-key notes as a proxy for sample quality.

In these experiments, we only use the fine-grained control in training, but do not insert any sampling control so that we can evaluate the inherent performance of the models themselves. The experiments were conducted across a range of guidance weights ($w$ from $0.5$ to $10$), and he results are summarized in Table~\ref{tab:cfg-weight-table}.

\begin{table}[ht]
\begin{center}
\renewcommand{\arraystretch}{1.2} 
\setlength{\tabcolsep}{10pt} 
\small 
\begin{tabular}{c|cccccc}
\hline
\rowcolor[HTML]{D9EAD3}  
\multicolumn{1}{c|}{\bf Values of $w$} &\multicolumn{1}{c}{\bf \% Out-of-Key} & \multicolumn{1}{c}{\bf OA} & \multicolumn{1}{c}{\bf OA} \\
\rowcolor[HTML]{D9EAD3}  
 & \bf Notes& \multicolumn{1}{c}{\bf (duration)} & \multicolumn{1}{c}{\bf (note density)} \\
\hline

\rowcolor[HTML]{FFFFFF}  
0.5 & 1.3\% &  $0.592$ & $0.803$\\
\rowcolor[HTML]{FFFFFF}  
& & $\pm 0.005$& $\pm 0.004$\\

\rowcolor[HTML]{F4F4F4}  
1.0 & 1.4\% &  $0.617$ & $0.830$\\
\rowcolor[HTML]{F4F4F4}  
&  & $\pm 0.005$& $\pm 0.003$\\

\rowcolor[HTML]{FFFFFF}  
3.0 & 1.7\% & $0.644$ & $0.848$\\
\rowcolor[HTML]{FFFFFF}  
& & $\pm 0.003$& $\pm 0.003$\\

\rowcolor[HTML]{F4F4F4}  
5.0 & 2.6\%  & $0.638$ & $0.846$\\
\rowcolor[HTML]{F4F4F4}  
& & $\pm 0.005$& $\pm 0.003$\\

\rowcolor[HTML]{FFFFFF}  
7.5 & 6.0\%  & $0.643$ & $0.829$\\
\rowcolor[HTML]{FFFFFF}  
&  & $\pm 0.005$& $\pm 0.004$\\

\rowcolor[HTML]{F4F4F4}  
10.0 & 14.3\% & $0.630$ & $0.779$\\
\rowcolor[HTML]{F4F4F4}  
& & $\pm 0.005$ & $\pm 0.005$\\

\hline
\end{tabular}
\end{center}
\caption{Comparison of the results with and without control in the sampling process.}
\label{tab:cfg-weight-table}
\end{table}
The findings indicate that as the guidance weight $w$ increases, the percentage of out-of-key notes rises, suggesting that lower $w$ values yield higher-quality samples. Meanwhile, the OA of duration and note density improves as $w$ increases from $0.5$ to $3.0$, indicating better alignment with rhythmic conditions. However, when $w$ exceeds $5.0$, a notable decline is observed in both the OA metrics and the percentage of out-of-key notes. This degradation is likely due to a significant drop in sample quality at excessively high $w$ values, where unnatural outputs undermine adherence to the rhythmic conditions. These observations are coherent with the existing results about the trade-off between sample quality and adherence to conditions in literature.

\section{Discussion}
The role of generative AI in music and art remains an intriguing question. While AI has demonstrated remarkable performance in fields such as image generation and language processing, these domains possess two characteristics that symbolic music lacks: an abundance of training data and well-designed objective metrics for evaluating quality. In contrast, for music, it is even unclear whether it is necessary to set the goal as generating compositions that closely resemble\footnote{or, in what sense?} some ``ground truth''. 

In this work, we apply fine-grained sampling control to eliminate out-of-key notes, ensuring that generated music adheres to the most common harmonies and chromatic progressions. This approach allows the model to consistently and efficiently produce music that is (in some ways) ``pleasing to the ear''. While suitable for the task of quickly creating large amounts of mediocre pieces, such models have a limited capability of replicating the artistry of a real composer, of creating sparkles with unexpected ``wrong'' keys by themselves. 


\end{document}